\documentclass[aip,jap,amsmath,amssymb,reprint,superscriptaddress,,longbibliography]{revtex4-1}
\usepackage{graphicx}
\usepackage{dcolumn}
\usepackage{bm}
\usepackage{color}
\begin{document}
\title{Spin coherence and depths of single nitrogen-vacancy centers created by ion implantation into diamond via screening masks}
\author{Shuntaro Ishizu}
\affiliation{School of Fundamental Science and Technology, Keio University, 3-14-1 Hiyoshi, Kohoku-ku, Yokohama 223-8522, Japan}
\author{Kento Sasaki}
\affiliation{School of Fundamental Science and Technology, Keio University, 3-14-1 Hiyoshi, Kohoku-ku, Yokohama 223-8522, Japan}
\author{Daiki Misonou}
\affiliation{School of Fundamental Science and Technology, Keio University, 3-14-1 Hiyoshi, Kohoku-ku, Yokohama 223-8522, Japan}
\author{Tokuyuki Teraji}
\affiliation{\mbox{National Institute for Materials Science, 1-1 Namiki, Tsukuba, Ibaraki 305-0044, Japan}}
\author{Kohei M. Itoh}
\email{kitoh@appi.keio.ac.jp}
\affiliation{School of Fundamental Science and Technology, Keio University, 3-14-1 Hiyoshi, Kohoku-ku, Yokohama 223-8522, Japan}
\affiliation{\mbox{Center for Spintronics Research Network, Keio University, 3-14-1 Hiyoshi, Kohoku-ku, Yokohama 223-8522, Japan}}
\author{Eisuke Abe}
\email{eisuke.abe@riken.jp}
\affiliation{School of Fundamental Science and Technology, Keio University, 3-14-1 Hiyoshi, Kohoku-ku, Yokohama 223-8522, Japan}
\affiliation{RIKEN Center for Emergent Matter Science, Wako, Saitama 351-0198, Japan}
\date{\today}
\begin{abstract}
We characterize single nitrogen-vacancy (NV) centers created by 10-keV N$^+$ ion implantation into diamond via thin SiO$_2$ layers working as screening masks.
Despite the relatively high acceleration energy compared with standard ones ($<$ 5~keV) used to create near-surface NV centers,
the screening masks modify the distribution of N$^+$ ions to be peaked at the diamond surface [Ito {\it et al.}, Appl.~Phys.~Lett.~{\bf 110}, 213105 (2017)].
We examine the relation between coherence times of the NV electronic spins and their depths,
demonstrating that a large portion of NV centers are located within 10~nm from the surface, consistent with Monte Carlo simulations.
The effect of the surface on the NV spin coherence time is evaluated through noise spectroscopy, surface topography, and X-ray photoelectron spectroscopy.
\end{abstract}
\maketitle
\section{Introduction}
The negatively charged nitrogen-vacancy (NV$^-$ or just NV) center in diamond has opened up a unique opportunity
for nanoscale nuclear magnetic resonance (NMR) spectroscopy and imaging.~\cite{MKS+13,SSP+13,MKC+14,HSR+15,DPL+15,LSU+16,APN+17}
NV's performance as a room temperature quantum sensor for nanoNMR is crucially dependent on
its depth from the diamond surface ($d_{\mathrm{NV}}$), as the NV center closer to the surface exhibits a stronger interaction with NMR analyte placed on the surface.
Furthermore, near-surface NV centers lead to efficient couplings with light,
which also adds impetus to other applications such as nanophotonics, plasmonics, and quantum communication.~\cite{TMO+13,SMZ+16,AEV+18,AHWZ18,WEH18,BGF+19}

There are several approaches to creating near-surface NV centers.
Most commonly used is low-energy N$^+$ ion implantation, typically with the acceleration energy less than 5~keV.~\cite{PNJ+10,OPC+12,dOAW+17,FBH+18,SDS+19}
Other methods such as nitrogen $\delta$-doping by chemical vapor deposition (CVD) and precise surface etching by plasma or thermal oxidation have also been demonstrated.~\cite{OHB+12,ORW+13,MDD+14,LPMD14,KMS+14,CGO+15,dOMW+15,ZZW+17}
Of particular interest to these platforms is the relation between $d_{\mathrm{NV}}$ and $T_2$, the coherence time of NV's electronic spin.
Longer $T_2$ is required to achieve better magnetic sensitivity, whereas near-surface NV centers tend to suffer short $T_2$.

In Ref.~\onlinecite{ISS+17}, some of the present authors reported that
near-surface NV centers can also be created by comparatively high energy (10~keV) N$^+$ ion implantation,
when combined with a SiO$_2$ screening mask deposited on the diamond surface.
The defining feature of this method is that the region of the highest N$^+$ density is located at the surface.
We showed this by Monte Carlo simulations, and experimentally confirmed the creation of NV centers
(see Sec.~\ref{sec_sample} for a more detailed description of our method and the result of Monte Carlo simulations).
The resulting N$^+$ distribution in our method is in contrast with the case for the standard (maskless) low-energy ion implantation,
in which the implanted N$^+$ ions are approximately Gaussian-distributed inside of the diamond.
An experimental confirmation of the distribution of the NV centers along the depth direction (depth profile),
as well as the relation between $T_2$ and $d_{\mathrm{NV}}$, are highly desired but have not yet been reported so far.
The present work addresses these issues.
Specifically, we focus on single NV centers and determine $d_{\mathrm{NV}}$ from NMR of proton ensembles on the surface, with direct application to nanoNMR in mind.~\cite{AS18,SIA18}

The rest of this paper is organized as follows.
In Sec.~\ref{sec_density}, we determine the density and yield of NV centers created by our method.
Section~\ref{sec_t2} presents the main result, the relation between $T_2$ and $d_{\mathrm{NV}}$,
and we analyze the depth profile of the NV centers by assembling the data from as many as 141 NV centers.
In Sec.~\ref{sec_noise}, we perform noise spectroscopy to investigate the source of the noise that limits $T_2$ in our sample.
The diamond surface is examined by atomic force microscopy (AFM) and X-ray photoelectron spectroscopy (XPS) in Sec.~\ref{sec_surface}.
We conclude in Sec.~\ref{sec_conclusion}.
To avoid digressing from the main subject, technical aspects and supporting data are presented in Appendices for interested readers.
They include the sample preparation procedure and the Monte Carlo simulations of N$^+$ ion implantation with screening masks (Sec.~\ref{sec_sample}), 
our experimental setup and microwave pulse sequences (Sec.~\ref{sec_pulse}),
and the result from double electron--electron resonance (DEER) measurements (Sec.~\ref{sec_deer}).
Although many of them are now becoming common tools among the researchers working on NV-based quantum sensing, it will still be beneficial to provide concise summaries.
We also refer to Ref.~\onlinecite{MSI+20} as a resource for further details about our experimental setup and techniques extensively used in the present work.

\section{Density and yield\label{sec_density}}
\begin{figure}
\begin{center}
\includegraphics{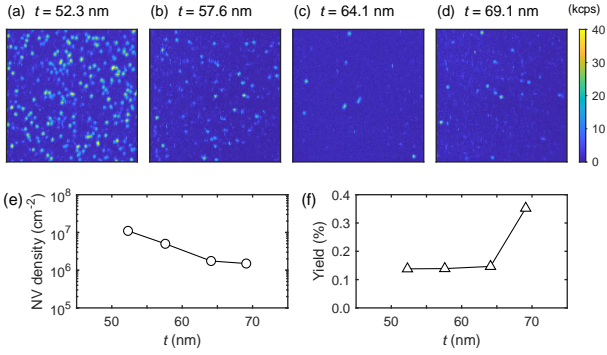}
\caption{(a--d) Fluorescence images (20~$\times$~20~$\mu$m$^2$) of diamond surfaces under the laser power of 0.6~mW.
$t$: thickness of the SiO$_2$ layer.
kcps: kilo-counts per second.
(e) NV density as a function of $t$.
(f) Yield as a function of $t$.
\label{fig1}}
\end{center}
\end{figure}

Figures~\ref{fig1}(a--d) show fluorescence images (20~$\times$~20~$\mu$m$^2$) of diamond surfaces,
where the SiO$_2$ layers of thickness $t$ = 52.3, 57.6, 64.1, and 69.1~nm were deposited prior to N$^+$ ion implantation.
It should be noted that these SiO$_2$ layers had been removed when these images were taken (see Sec.~\ref{sec_sample}).
Each image shows multiple bright spots, the number of which is decreasing as $t$ increases.
To confirm the presence of single NV centers, continuous wave optically detected magnetic resonance (CW ODMR) was performed on individual bright spots.
For the areas with $t$ = 64.1 and 69.1~nm, all the bright spots in the searched area were examined and the number of ODMR-active points were counted.
The areal density was then calculated by dividing it by the searched area.
For the $t$ = 52.3 and 57.6~nm areas, where many more bright spots exist, we set a threshold count rate, above which we judged as an emission from a single NV center,
and the number of above-threshold spots were counted by a computer program.
The threshold count rates were 2.0~$\times$~10$^4$~kcps for $t$ = 52.3~nm and 1.7~$\times$10$^4$~kcps for $t$ = 57.6~nm,
which were the minimum photon count rates among a subset of the bright spots on which CW ODMR was performed and the signals were observed.
The result is summarized in Fig.~\ref{fig1}(e).

We also calculate the yield (efficiency of conversion from N$^+$ ions into NV$^-$ centers) as shown in Fig.~\ref{fig1}(f).
It is given by the ratio of the areal density divided by the implanted N$^+$ density deduced from Monte Carlo simulations.
The low yield of 0.1--0.4\% is consistent with our previous observation,~\cite{ISS+17}
and similar to the values obtained in low energy ion implantation ($\approx$ 3~keV).~\cite{PNJ+10}
The cause of the latter case is attributed to the lack of vacancies to pair up with nitrogen atoms to form NV centers,
but it has been shown that the yield can be improved by carefully adjusting annealing, chemical or plasma etching conditions.~\cite{OSP+13,YUW+13,AHA+14}
This aspect has not been optimized yet in our case, and can be improved in the future.

\section{Coherence time and depth\label{sec_t2}}
\begin{figure}
\begin{center}
\includegraphics{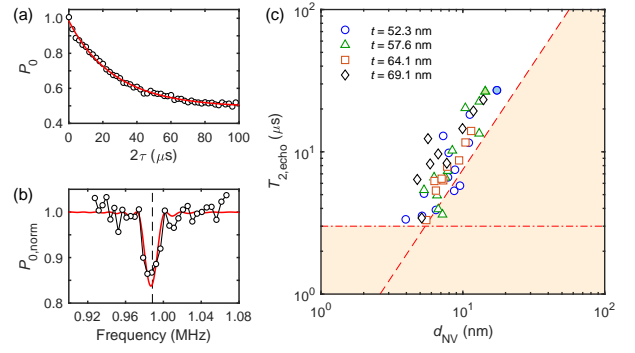}
\caption{(a) Hahn echo decay.
(b) NMR spectrum.
(c) Relation between $T_{2,\mathrm{echo}}$ and $d_{\mathrm{NV}}$.
The dashed line is the measurement limit as determined by Eq.~(\ref{eq_t2limit}), whereas the dash--dot line indicates $T_2$ = 3~$\mu$s.
They define the region (unpainted) that is ``accessible'' in our experiments.
See the main text for detail.
The filled circle indicates the NV center with which the Hahn echo decay in (a), the NMR spectrum in (b), and the noise spectroscopy data in Fig.~\ref{fig4} were taken.
The filled triangle indicates the NV center with which the DEER data in Fig.~\ref{fig8} were taken.
\label{fig2}}
\end{center}
\end{figure}

Figure~\ref{fig2}(a) shows an example of Hahn echo decay [see Fig.~\ref{fig7}(a) for the pulse sequence].~\cite{H50}
The static magnetic field $B_0$ = 23.2~mT was applied parallel to the NV symmetry axis,
and the microwave frequency was set at $\omega_{0,-1}/2\pi$ = 2218.2~MHz, corresponding to the $m_S = 0 \leftrightarrow -1$ transition.
$P_0$ is the probability of the final spin state being $m_S$ = 0, which we calculate from calibrated photon counts.~\cite{MSI+20}
We describe a general coherence decay curve as
\begin{equation}
P_0 = \frac{1}{2} + \frac{ \mathcal{C} }{2} \exp \left[ -\left( \frac{ t_{\mathrm{tot}} }{T_2} \right)^p \right].
\label{eq_t2}
\end{equation}
The form of $\mathcal{C}$ depends on a type of measurements, $t_{\mathrm{tot}}$ is the total time of the pulse sequence applied.
Fitting parameters are $T_2$ and $p$ (stretched exponent).
In the case of Hahn echo, $\mathcal{C}_{\mathrm{echo}}$ = 1 and $t_{\mathrm{tot}}$ = 2$\tau$.
The solid line in Fig.~\ref{fig2}(a) is the fit, giving $T_{2,\mathrm{echo}}$ = 27.1~$\mu$s and $p$ = 0.98.

We then apply a multipulse sequence to determine $d_{\mathrm{NV}}$ from NMR of protons on the surface.
We use XY4, XY8, and XY16 sequences (see Sec.~\ref{sec_pulse} for the definitions).~\cite{GBC90}
When the total number of $\pi$ pulses in a sequence is $N$, it is denoted as XY$k$-$N$ ($k$ = 4, 8, or 16 and $N$ is an integer multiple of $k$).
The decay is given by setting $t_{\mathrm{tot}}$ = $N\tau$, and the NMR spectrum is described by~\cite{PDC+16} 
\begin{eqnarray}
& & \mathcal{C}_{\mathrm{NMR}} = \\
& & \exp \left\{ - 2 \left( \frac{ \gamma_{\mathrm{e}} B_{\mathrm{rms}} N\tau }{ \pi } \right)^2 
\mathrm{sinc}^2 \left[ \frac{N \tau}{2} \left( \omega_{\mathrm{n}} - \frac{\pi}{\tau} \right) \right] \right\}. \nonumber
\end{eqnarray}
Here, $\gamma_{\mathrm{e}}/2\pi$ = 28~MHz/mT is the gyromagnetic ratio of the NV spin, and
$\omega_{\mathrm{n}}/2\pi = \gamma_{\mathrm{h}} B_0/2\pi$ is the proton NMR frequency
with $\gamma_{\mathrm{h}}/2\pi$ = 42.577~kHz/mT the gyromagnetic ratio of the $^1$H nuclear spin.
$B_{\mathrm{rms}}$ is the root-mean-square nuclear dipolar magnetic field integrated over the semi-infinite volume (i.e., outside of the diamond),
and can be calculated analytically as~\cite{SSP+13,LPMD14,RDO+14,PDC+16} 
\begin{equation}
B_{\mathrm{rms}} = \frac{\mu_0 \hbar \gamma_{\mathrm{h}} }{4 \pi} \sqrt{ \frac{ 5 \pi \rho}{ 96 \, d_{\mathrm{NV}}^3 } }
\end{equation}
with $\mu_0$ the permeability of vacuum, $\hbar$ the reduced Planck constant, and
$\rho$ the proton density of the immersion oil on the surface, known to be 6~$\times$~10$^{28}$~m$^{-3}$ in the present case (Olympus IMMOIL-F30CC).

Figure~\ref{fig2}(b) is the NMR spectrum, taken with the NV center of Fig.~\ref{fig2}(a) and the XY16-128 sequence.
The stretched exponential decay component is subtracted (normalized $P_0$, or $P_{0,\mathrm{norm}}$),
and the horizontal axis is $(2\tau)^{-1}$ in the frequency unit.
The NMR dip appears at 0.988~MHz, agreeing with the calculated $\omega_{\mathrm{n}}/2\pi$. 
From the fit, $d_{\mathrm{NV}}$ is determined to be 17.4~nm.
(Consult Fig.~10 of Ref.~\onlinecite{MSI+20} for another example of the NMR spectrum measured with a different, shallower NV center in the same sample.)

We repeat the measurements on different NV centers, and the relation between $T_{2,\mathrm{echo}}$ and $d_{\mathrm{NV}}$,
obtained from 43 NV centers in total, is plotted in Fig.~\ref{fig2}(c).
[The point with the filled circle corresponds to the NV center examined in Figs.~\ref{fig2}(a,b).]
We note again that, although the NV centers are grouped by $t$ (SiO$_2$ thickness), the SiO$_2$ layer had been removed when these measurements were performed.
A general trend is clearly discerned: the deeper the NV center, the longer the coherence,
as observed in previous works using different methods to create NV centers.~\cite{dOAW+17,FBH+18,SDS+19}
For instance, we find the observed $T_{2,\mathrm{echo}}$--$d_{\mathrm{NV}}$ relation is quantitatively close to that obtained in Ref.~\onlinecite{SDS+19} (without high temperature annealing).
It must be emphasized, however, that in our case a large fraction of NV centers exhibited short $T_2$ insufficient to observe clear NMR spectra.
Consequently, we could not determine their depths, and such NV centers are not shown in Fig.~\ref{fig2}(c).
We will include them when estimating the depth profile,
but before doing so we discuss an experimental condition to detect NMR and set the region in the $T_{2,\mathrm{echo}}$--$d_{\mathrm{NV}}$ space
where the experimental observation can be made.

The condition is simply that the signal $\mathcal{S}$ be larger than the noise $\mathcal{N}$.
The NMR signal arises from the component of $P_0$ after subtracting the coherence decay,
therefore $\mathcal{S} = \frac{1}{2} (1-\mathcal{C}_{\mathrm{NMR}}) e^{-(N\tau/T_2)^p}$.
We set $p$ = 1 for simplicity (also consistent with experimental observation) and $\tau = \pi/\omega_{\mathrm{n}}$ = 506.2~ns, corresponding to the resonance condition.
Experimentally, the number of $\pi$ pulses $N$ is determined from the balance between the signal strength and the coherence time,
and we usually found $N\tau/T_{2,\mathrm{echo}} \approx 2$ to be an appropriate operating condition.
For instance, the condition for Fig.~\ref{fig2}(b) is $N\tau/T_{2,\mathrm{echo}}$ = 128~$\times$~506.2~ns/27.1~$\mu$s $\approx$ 2.4.
Here, $T_{2,\mathrm{echo}}$ is used, as we do not know $T_2$ under dynamical decoupling in advance.

The measurement noise $\mathcal{N}$ arises from the Poisson distribution of the photon counts $n$, the variance of which is given by $\sqrt{n}$.
The signal photon count $n_{\mathrm{s}}$ is defined as $n_{\mathrm{s}} = n_1 + (n_0 - n_1) P_0$,
where $n_0$ and $n_1$ are the photon counts from the $m_S$ = 0 and $-$1 states, respectively, and are mutually related by $n_1 = (1-c) n_0$, with $c$ the contrast.
We obtain $P_0 = (n_{\mathrm{s}} - n_1)/(c n_0) $
and the fluctuation of $n_{\mathrm{s}}$, given by $\sqrt{n_{\mathrm{s}}}$, leads to $\mathcal{N} = \sqrt{ n_{\mathrm{s}} }/(c n_0) \approx (c \sqrt{n_0})^{-1}$.
Combining these, we write the condition $\mathcal{S}/\mathcal{N} = 1$ as
\begin{equation}
T_{2,\mathrm{echo}} = \frac{ \pi }{ 2\sqrt{2} \gamma_e B_{\mathrm{rms}} } \sqrt{-\ln \left( 1- \frac{2 e^2}{ c \sqrt{n_0} } \right) }.
\label{eq_t2limit}
\end{equation}
Typical values for $c$ and $n_0$ in the present experimental condition are $c \approx$ 0.2 and $n_0 \approx 5 \times 10^5$ counts.
Substituting these values, we draw the dashed line in Fig.~\ref{fig2}(c), below which the depth cannot be determined.
In addition to this constraint, short $T_{2,\mathrm{echo}}$ limits the number of pulses we can apply.
For instance, to satisfy $N\tau/T_{2,\mathrm{echo}} \approx 2$ with $N$ = 8, $T_{2,\mathrm{echo}}$ needs to be longer than 2~$\mu$s.
In fact, from NV centers with $T_{2,\mathrm{echo}}$ shorter than 3~$\mu$s we were not able to observe clear NMR signals
(and the extension of $T_2$ by dynamical decoupling is often not effective),
even though they are likely to be located close to the surface.
This motivates us to draw another bound $T_{2,\mathrm{echo}}$ = 3~$\mu$s in Fig.~\ref{fig2}(c).
Although the argument here is manifestly crude, the defined ``accessible'' region (unpainted) reasonably coincides with the experimental observation.

\begin{figure}
\begin{center}
\includegraphics{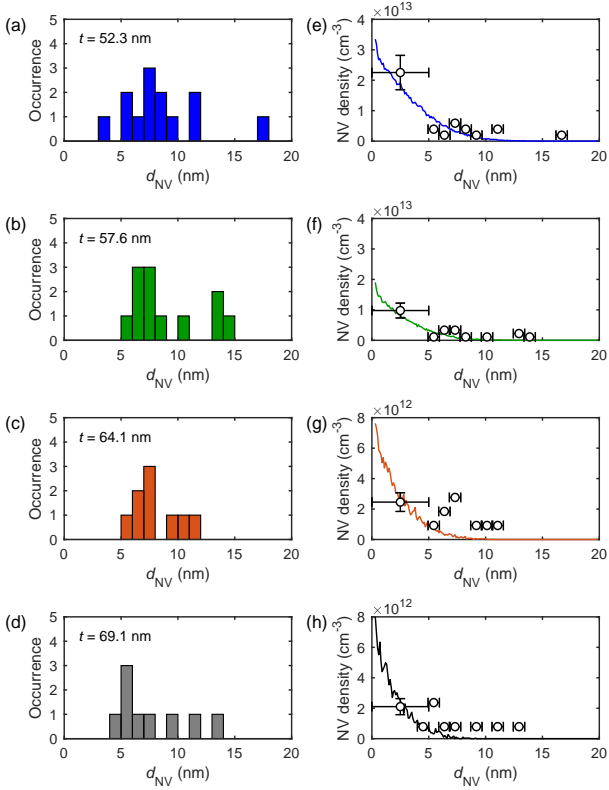}
\caption{(a--d) Histograms showing the distributions of the 43 {\it depth-determined} NV centers.
(e--h) NV densities estimated from the experiments ($\bigcirc$), 98 {\it depth-undetermined} NV centers included, and the simulations (solid lines).
See the main text for detail.
\label{fig3}}
\end{center}
\end{figure}

From the dataset in Fig.~\ref{fig2}(c), it is straightforward to produce histograms
showing the number of NV centers (occurrence) as a function of $d_{\mathrm{NV}}$ [Figs.~\ref{fig3}(a--d)].
Ideally, these histograms are to be directly compared with the depth profile from Monte Carlo simulations (Fig.~\ref{fig6}).
However, as mentioned above, the fact that the data points in Fig.~\ref{fig2}(c) are limited to a certain region suggests
that they alone are not sufficient to deduce a firm conclusion about the depth profile.
There are indeed another 98 NV centers we measured but $d_{\mathrm{NV}}$ were not determined.
This number breaks down into 44 ($t$ = 52.3~nm), 33 (57.6~nm), 10 (64.1~nm), and 11 (69.1~nm),
not negligible compared with the numbers of the depth-determined NV centers, 13 (52.3~nm), 12 (57.6~nm), 9 (64.1~nm), and 9 (69.1~nm).

To include the former group in the histograms, we assume, judging from the trend observed in Fig.~\ref{fig2}(c), that they are located at $d_{\mathrm{NV}} \leq$ 5~nm.
In Figs.~\ref{fig3}(e--h), the solid lines are the Monte Carlo simulations in Fig.~\ref{fig6} multiplied by the respective yields in Fig.~\ref{fig1}(f).
The circle points ($\bigcirc$) for $d_{\mathrm{NV}} >$ 5~nm are the histograms multiplied by the respective NV densities.
Similarly, those for $d_{\mathrm{NV}} \leq$ 5~nm are based on the numbers of the NV centers whose depths left undetermined, but divided by five to reflect the fact that
we are agnostic on which ``bin'' (the width of 1~nm) they are in.
Thus, the horizontal error bar for them is set as $d_{\mathrm{NV}}$ = 2.5~$\pm$ 2.5~nm (the other error bars are set as the bin width).
From Figs.~\ref{fig3}(e--h), it is interpreted, though not conclusive, that the profile is concentrated toward the surface,
unlike the standard ion implantation without a screening mask.

\section{Noise spectroscopy\label{sec_noise}}
In our sample, it is observed that the NV centers with short $T_{2,\mathrm{echo}}$ also tend to be short-lived, degrading $T_{2,\mathrm{echo}}$ during the experiments.
In one case, initially $T_{2,\mathrm{echo}}$ was 6~$\mu$s, but it reduced continuously to 2~$\mu$s in 5 hours, then fluctuates between 1 and 2~$\mu$s.
Another case shows that initial $T_{2,\mathrm{echo}}$ of 18~$\mu$s slowly degrades to 12~$\mu$s in the time scale of 80~hours.
These are likely to arise from the coupling with the surface, and we next examine the origin of the noise that affects the NV spin.
For this purpose, we would ideally hope to use shallower NV centers, but as mentioned above, they suffer degradation of $T_{2,\mathrm{echo}}$,
making a clean set of experiments difficult.
Nonetheless, the trend in Fig.~\ref{fig2}(c) indicates that the plotted NV centers share the same noise source.
We therefore chose to use the NV center with longest $T_{2,\mathrm{echo}}$ found in the $t$ = 52.3~nm area [the filled circle in Fig.~\ref{fig2}(c)].
For the noise spectroscopy, $B_0$ was fixed at 23.2~mT, where $T_2$ and $d_{\mathrm{NV}}$ had also been measured.

\begin{figure}
\begin{center}
\includegraphics{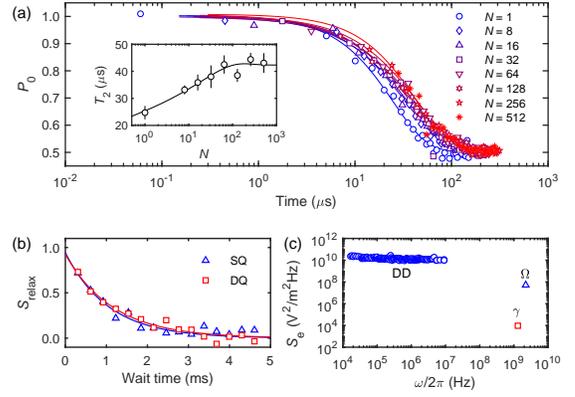}
\caption{(a) Coherence decay curves under dynamical decoupling.
The $N$ = 1 data were taken by the Hahn echo (the time axis is given by 2$\tau$),
the $N$ = 8 data by XY8, and the $N \geq$ 16 data by XY16-$N$ (the time axis is given by $N\tau$).
The inset: $T_2$ as a function of $N$.
The horizontal bars indicate the 95\% confidence limits.
(b) Relaxation signals for the SQ ($\triangle$) and DQ ($\square$) channels.
(c) Noise spectrum $S_{\mathrm{e}}(\omega)$ deduced from the spectral decomposition of the dynamical decoupling data ($\bigcirc$),
the SQ relaxation rate $\Omega$ ($\triangle$), and the DQ relaxation rate $\gamma$ ($\square$).
All the measurements were performed at $B_0$ = 23.2~mT.
\label{fig4}}
\end{center}
\end{figure}

We first perform dynamical decoupling by increasing $N$ up to 512.
The decay curves are shown in Fig.~\ref{fig4}(a).
Although dynamical decoupling improves the coherence to some extent, the effect is not significant.
$T_2^{(N)}$, $T_2$ by $N$-pulse dynamical decoupling, shows a scaling and saturating behavior described by
\begin{equation}
T_2^{(N)} = T_2^{(1)} [ N_{\mathrm{sat}}^s + (N^s - N_{\mathrm{sat}}^s) e^{-N/N_{\mathrm{sat}}} ],
\end{equation}
and the fit gives $N_{\mathrm{sat}}$ = 105.2, $s$ = 0.10, and $T_2^{(\infty)}$ = $T_2^{(1)} N_{\mathrm{sat}}^{s}$ = 43.0~$\mu$s [the inset of Fig.~\ref{fig4}(a)].
$T_2^{(1)}$ corresponds to $T_{2,\mathrm{echo}}$, but is treated as a parameter here.
The scaling exponent $s$ of 2/3 is expected for an environment that fluctuates slowly compared with $T_2$.~\cite{BPB+12,RMU+15}
The small $s \ll$ 2/3 as well as the saturated $T_2$ observed here indicate the presence of broadband noise.

This observation is reinforced by spectral decomposition.
Under a few simplifying assumptions, the spectral component at $\omega = \pi/\tau$ is given by~\cite{BPB+12,RMU+15}
\begin{equation}
S_{\mathrm{DD}}(\omega) = -\frac{ \pi \ln(2P_0 -1)}{ N \tau }.
\end{equation}
One can compute the noise spectrum directly from the dynamical decoupling data.
Figure~\ref{fig4}(c) shows the spectral function $S_{\mathrm{e}}(\omega)$,
related to $S_{\mathrm{DD}}(\omega)$ by $S_{\mathrm{e}}(\omega) = 2S_{\mathrm{DD}}(\omega)/d_{\parallel}^2$,
where $d_{\parallel}$ = 0.35~Hz~cm/V is the electric dipole moment parallel to the NV symmetry axis.~\cite{MAJ17}
The reason for this conversion is to enable a comparison between the (effective) electric and magnetic field noises.
Note that in Fig~\ref{fig4}(c) the data points close to $(2P_0 -1)$ = 0 or 1 that are insensitive to the change in $\tau$ have been removed.
Further analysis usually invokes a Lorentzian fit~\cite{BPB+12,RMU+15}
\begin{equation}
S_{\mathrm{DD}}(\omega) = \frac{ \langle E_{\perp}^2 \rangle \tau_{\mathrm{c}} }{ \pi [1+(\omega \tau_{\mathrm{c}})^2] } 
\end{equation}
or a sum of multiple Lorentzians, 
where $\langle E_{\perp}^2 \rangle$ is the average coupling strength of the environment to the NV center,
and $\tau_{\mathrm{c}}$ is the correlation time of the environment.
In the present case, the spectrum is almost flat (typically 1--2$\times$10$^{10}$~V$^2$/m$^2$Hz)
up to $\omega/2\pi \approx$ 10$^7$~Hz in the double logarithmic plot,
suggesting that the noise is white in the probed frequency range ($\omega \tau_{ \mathrm{c}} \ll 1$).
This makes fitting to a Lorentzian curve arbitrary, and we can only infer $\tau_{ \mathrm{c}}$ to be shorter than 100~ns.
Still, the short $\tau_{ \mathrm{c}} \ll T_2$ is consistent with the small $s$.

We also probe the noise by relaxation measurements through single quantum (SQ) and double quantum (DQ) channels
as shown in Fig.~\ref{fig4}(b) [see Fig.~\ref{fig7}(c) for the pulse sequence].~\cite{MAJ17}
The relaxation times of the two channels are almost the same,
$T_{1,\mathrm{SQ}}$ = 1.06~ms and $T_{1,\mathrm{DQ}}$ = 1.16~ms,
consistent with the previous observations at similar magnetic fields.~\cite{MAJ17,SDS+19,GCK20}
We can then deduce the SQ and DQ relaxation rates, $\Omega$ and $\gamma$,
using the relations $T_{1,\mathrm{SQ}} = (3 \Omega)^{-1}$ and $T_{1,\mathrm{DQ}} = (\Omega+2\gamma)^{-1}$.~\cite{MAJ17}
We obtain $\Omega$ = 316~Hz, corresponding to the magnetic noise probed at $\omega_{0,-1}/2\pi$ = 2218.2~MHz,
and $\gamma$ = 272~Hz, corresponding to the electric noise probed at $\omega_{-1,1}/2\pi$ = 1301.5~MHz (the separation of the $m_S$ = $-$1 and 1 states at 23.2~mT).
$\Omega$ and $\gamma$ are converted into the spectral components as
$2 \Omega/d_{\parallel}^2$ = 5.16$\times$10$^7$~V$^2$/m$^2$Hz and
$\gamma/d_{\perp}^2$ = 9.41$\times$10$^{3}$~V$^2$/m$^2$Hz, respectively,
with $d_{\perp}$ = 17~Hz~cm/V the electric dipole moment perpendicular to the NV symmetry axis.~\cite{MAJ17,SDS+19}
The combined plot in Fig.~\ref{fig4}(c) certifies that in the wide frequency region the dominant contribution comes from the magnetic noise,
which is presumably due to surface defects, as has been argued in the case of low energy ion implantations.~\cite{RMU+15,MAJ17,SDS+19}

\section{Surface condition\label{sec_surface}}
We have found that our method of creating near-surface NV centers provides the $T_2$--$d_{\mathrm{NV}}$ relation comparable to other existing methods.
Yet, the SiO$_2$ deposition, in which an amorphous SiO$_2$ source (melting point $\approx$ 1700$^{\circ}$C) is heated by an electron beam
and sublimated SiO$_2$ is directed to the diamond substrate in vacuum,
is the process unique to ours, and it is of interest to examine an effect it might have on the diamond surface.
Here, we perform AFM and XPS on two areas, one in which SiO$_2$ was not deposited and ion implantation was blocked (denoted as $t$ = 0),
and the other in which ion implantation was performed through SiO$_2$ layers (denoted as $t \neq$ 0).
These measurements were performed after all the ODMR-based experiments were done (including DEER presented in Sec.~\ref{sec_deer}),
and therefore the $t$ = 0 area was subject to the same chemical and annealing processes (described in Sec.~\ref{sec_sample}) as the $t \neq$ 0 area.

\begin{figure}
\begin{center}
\includegraphics{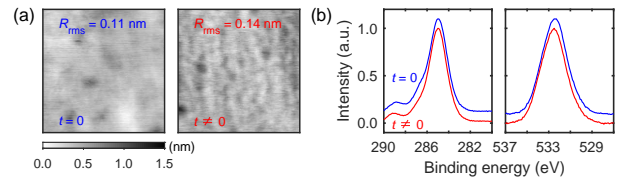}
\caption{(a) AFM topographic images (500~$\times$~500~nm$^2$) of the diamond surfaces.
(b) The normalized XPS spectra showing the carbon 1$s$ (left) and oxygen 1$s$ (right) peaks.
The spectra for $t \neq$ 0 are shifted by 0.1 for clarity. 
\label{fig5}}
\end{center}
\end{figure}

Figure~\ref{fig5}(a) compares AFM topographic images (500~$\times$~500~nm$^2$) taken at the $t$ = 0 and $t \neq$ 0 areas.
To be more precise, the latter belongs to the $t$ = 52.3~nm area, but is not correlated with the area on which the ODMR experiments were carried out.
In these particular areas shown in Fig.~\ref{fig5}(a), the root-mean-square roughness, $R_{\mathrm{rms}}$, is 0.11~nm for $t$ = 0 and 0.14~nm for $t \neq$ 0.
The measurements were repeated on different locations, and $R_{\mathrm{rms}}$ averaged over the measured areas were 0.17~nm for both $t$ = 0 and $t \neq$ 0,
indicating that the SiO$_2$ deposition and subsequent ion implantation have not altered the surface condition significantly.
Nonetheless, the change in the surface morphology, with more `wrinkles' in the latter, is discernible.
It is not clear yet whether these wrinkles affect the coherence properties of the NV spins.

Figure~\ref{fig5}(b) shows XPS spectra of carbon 1$s$ and oxygen 1$s$.
The C 1$s$ peak at 285~eV arises primarily from $sp^3$ (in bulk),
and the asymmetric shape indicates the presence of other bonds such as
C--O, C=O ($>$ 285~eV), and $sp^2$ ($<$ 285~eV).~\cite{BDM+14,SDS+19}
The O 1$s$ peak at 532~eV has a broad linewidth, presumably composed of 
C--O--C, C--C--H, and C=O ($<$ 532~eV).~\cite{BDM+14,SDS+19}
Despite the limited spectral resolution, we can conclude that the chemical composition of the surface is essentially the same between the $t$ = 0 and $t \neq$ 0 areas.

\section{Conclusion and outlook\label{sec_conclusion}}
To conclude, we created single nitrogen-vacancy (NV) centers by 10-keV N$^+$ ion implantation into diamond via SiO$_2$ layers of $t$ = 52.3, 57.6, 64.1, and 69.1~nm,
and examined the relation between $T_{2,\mathrm{echo}}$ and $d_{\mathrm{NV}}$.
The depth profiles incorporating the NV centers with $T_2 <$ 3~$\mu$s were found to be consistent with Monte Carlo simulations.
The effect of the surface on $T_2$ was examined by a combination of noise spectroscopy, surface topography, and XPS.

Although our method is found to give the $T_2$--$d_{\mathrm{NV}}$ relation comparable to other existing methods,
we also conceive that the surface undergoes a slight change in morphology due to SiO$_2$ deposition and $T_2$ becomes unstable and/or is shortened over time.
Recently, these aspects have been investigated in great detail by Sangtawesin {\it et al.},~\cite{SDS+19}
who emphasized the importance of keeping the surface purity and the quality of surface morphology in each step of sample preparation.
Longer coherence was achieved by combining annealing at 800$^{\circ}$C, at 1200$^{\circ}$C, and in oxygen atmosphere,
which are thought to remove residual vacancy complexes due to ion implantation and lead to highly ordered oxygen terminated surface.
While we did not adopt annealing at 1200$^{\circ}$C, this additional process is reported to improve $T_2$ several times longer (e.g., a few $\mu$s to ten $\mu$s)
and achieve $R_{\mathrm{rms}}$ of 0.06~nm (compared with our $R_{\mathrm{rms}}$ = 0.17~nm).
Given that without 1200$^{\circ}$C the $T_2$--$d_{\mathrm{NV}}$ relation in the present work and that in Ref.~\onlinecite{SDS+19} are similar,
we expect the process to provide an important clue for stabilizing and extending $T_2$ of NV center within 5~nm from the surface.

\section*{Acknowledgement}
K.S. was supported by the JSPS Grant-in-Aid for Research Fellowship for Young Scientists (DC1), Grant No.~JP17J05890.
T.T. was supported by the JSPS Grant-in-Aid for Scientific Research (KAKENHI) (B) Grant No.~20H02187 and (B) Grant No.~19H02617,
(S) Grant No.~16H06326
the JST CREST (JPMJCR1773), and MEXT Q‐LEAP (JPMXS0118068379).
K.M.I. was supported by the JSPS KAKENHI (S) Grant No.~26220602 and (B) Grant No.~19H02547,
the JST Development of Systems and Technologies for Advanced Measurement and Analysis (SENTAN),
and the Spintronics Research Network of Japan (Spin-RNJ).
The authors thank the technical staff of Central Service Facilities for Research at Keio University
for their assistance in XPS measurements.
The data that support the findings of the present work are available from the corresponding author upon reasonable request.

\appendix
\section{Monte Carlo simulation and sample preparation\label{sec_sample}}
\begin{figure}
\begin{center}
\includegraphics{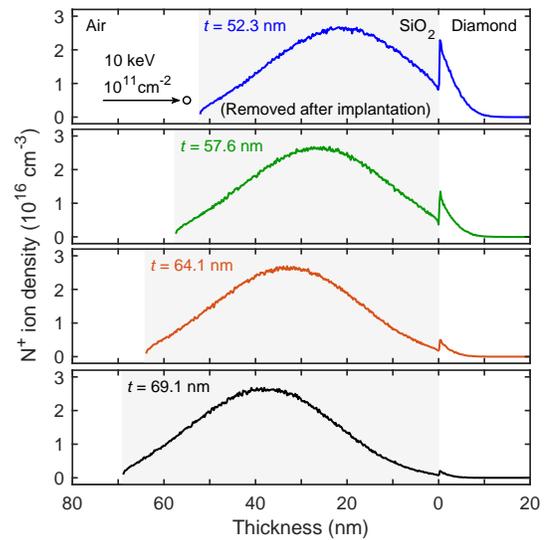}
\caption{Monte Carlo simulations of N$^+$ ion implantations into diamond with SiO$_2$ layer of thickness $t$ = 52.3, 57.6, 64.1, and 69.1~nm on top.
The acceleration energy is 10~keV and the dose is 10$^{11}$~cm$^{-2}$.
The incident angle is normal to the surface.
\label{fig6}}
\end{center}
\end{figure}

Figure~\ref{fig6} shows Monte Carlo simulations of N$^+$ ion implantation into diamond through SiO$_2$ layers,
using a software package SRIM (Stopping and Range of Ions in Matter).~\cite{SRIM,Z04}
Some merits of using a screening mask are immediately apparent;
once the SiO$_2$ layer is removed after ion implantation, the region of the highest NV density is located {\it at the surface},
with the distributions in the depth direction much narrower than those without SiO$_2$.
It is also clear that the resulting N$^+$ density is significantly reduced from that expected without a screening mask.
With the dose of 10$^{11}$~cm$^2$ and N$^+$-to-NV$^-$ conversion efficiency (yield) of the order of 0.1~\%, as observed in the present work,
the resulting NV density will be low enough for single NV centers to be resolved optically. 
Less clear from Fig.~\ref{fig6} is that the presence of the amorphous SiO$_2$ layer mitigates the ion channeling.
SRIM does not take into account the ion channeling and therefore the actual depth profile for a crystalline material such as diamond 
will have a longer tail toward the interior of the material.
In the present case, the simulated profiles should be more reliable.
Consult Ref.~\onlinecite{ISS+17} for simulation results covering wider ranges of $t$ and additional analysis.

The basic procedures for sample preparation are as described in Ref.~\onlinecite{ISS+17}, except for two points.
The first is the growth of isotopically pure $^{12}$C diamond layer, and the second is the use of $^{14}$N as an implantation species, instead of $^{15}$N. 
We began with a natural abundant (1.1\% of $^{13}$C), type-IIa (001) diamond substrate from Element Six.
The size of the substrate was 2~$\times$~2~$\times$~0.5~mm$^{3}$.
On top of it, an undoped, $^{12}$C (99.95~\%) layer was grown by CVD with the thickness of a few micron.~\cite{T15}
Prior to ion implantation, thorough search of NV centers were made, but none was found from the CVD-grown layer. 
It may seem that the natural abundant substrate was sufficient,
as the observed $T_2$ of the NV spins were short ($<$ 30~$\mu$s) compared with the $T_2$ timescale set by the nuclear spin bath.
Nonetheless, the use of $^{12}$C suppresses ``collapse and revival'' of the Hahn echo signal, facilitating both experiments and data analysis.~\cite{CDT+06}
It also avoids misinterpretation of the NMR signal due to the forth spurious harmonics of nearby $^{13}$C nuclei overlapping with the ensemble proton signal.~\cite{LBR+15}

Electron beam evaporation was used to create SiO$_2$ layers with controlled thickness.
A metal plate with four apertures with the diameters of 500~$\mu$m was placed on the diamond surface,
and in each evaporation run, one aperture was open.
The thickness of the SiO$_2$ layer was monitored by placing a silicon substrate in the vicinity of the sample,
and ellipsometry was performed on the deposited silicon substrate.
After four evaporation runs, the sample was implanted with $^{14}$N$^+$ ions at 10~keV and the dose of 10$^{11}$~cm$^{-2}$ (Ion Technology Center), with all the apertures open.
Previously, $^{15}$N isotopes were used to discriminate from the $^{14}$N isotopes in the bulk (99.6~\%),
because in the previous work we did not check the depths of the NV centers.
This time, the presence of an undoped CVD layer and the thorough analysis of the NV depth rendered the use of $^{15}$N unnecessary.
After ion implantation, the SiO$_2$ layers were removed by hydrofluoric acid.
The sample was subsequently annealed at 800~$^{\circ}$C for 2~hours in vacuum (9.7~$\times$~10$^{-7}$~torr) in order to let vacancies diffuse to form NV centers
and at 450~$^{\circ}$C for 9~hours in oxygen atmosphere in order to convert neutral NV (NV$^0$) centers into negatively charged ones (NV$^-$).~\cite{FSB+10}
The sample was chemically cleaned after CVD and ion implantation by a 1:1:1 mixture of concentrated sulfuric, nitric, and perchloric acids (triacids),
and after annealing by a 1:2 mixture of hydrogen peroxide and concentrated sulfuric acid (piranha).

\section{Experimental setup and pulse sequence\label{sec_pulse}}
The experimental setup used in the present work, including our method to generate microwave pulses, is described in detail in Ref.~\onlinecite{MSI+20}.
It is a homebuilt tabletop scanning confocal microscope combined with high-frequency electronics,
enabling time-domain and multifrequency ODMR of single NV centers.
All the sequences start and end by green (532~nm) laser illumination for initialization (Init.) and readout (RO).
The photons in the range between 650 and 800~nm are detected by a single-photon counting module.
To deliver microwave to the sample, broadband large-area microwave antennas were used in standard ODMR and proton NMR.~\cite{SMS+16}
For relaxation and DEER experiments, which also require frequencies outside of the bandwidth of our microwave antennas, a copper wire was used.

\begin{figure}
\begin{center}
\includegraphics{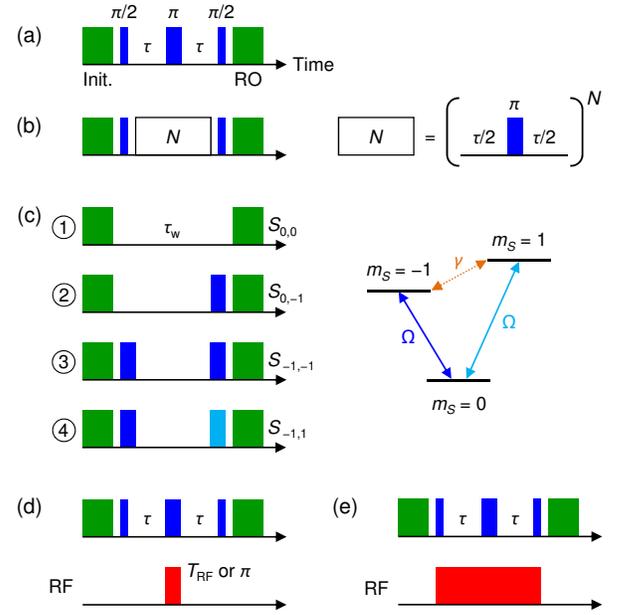}
\caption{Summary of microwave pulse sequences.
(a) Hahn echo sequence.
The green boxes indicate the laser illumination.
The narrow and wide boxes indicate the microwave $\pi/2$ and $\pi$ pulses, respectively.
The box color indicates the transition used, in this case the $m_S = 0 \leftrightarrow -1$ transition (see the energy level diagram). 
(b) Dynamical decoupling sequence.
(c) Relaxation measurements.
\textcircled{\scriptsize 1} (\textcircled{\scriptsize 2}) The NV spin is in the $m_S$ = 0 state
during the wait time $\tau_{\mathrm{w}}$ and the population of the $m_S$ = 0 ($m_S$ = $-$1) state is measured, giving the signal $S_{0,0}$ ($S_{0,-1}$).
\textcircled{\scriptsize 3} (\textcircled{\scriptsize 4}) The NV spin is in the $m_S$ = $-$1 state during $\tau_{\mathrm{w}}$ and the population of the $m_S$ = $-$1 ($m_S$ = 1) state is measured, giving the signal $S_{-1,-1}$ ($S_{-1,1}$).
$S_{0,0} - S_{0,-1}$ ($S_{-1,-1} - S_{-1,1}$) gives the relaxation through the SQ (DQ) channel.
(d) DEER sequence.
When the RF pulse length $T_{\mathrm{RF}}$ is varied, the Rabi oscillation of the resonant paramagnetic defect is measured.
By setting $T_{\mathrm{RF}}$ at an approximate $\pi$ pulse length and changing the RF frequency, the DEER spectrum is obtained. 
(e) Continuous driving.
\label{fig7}}
\end{center}
\end{figure}

Microwave pulse sequences used in the present work are summarized in Fig.~\ref{fig7}. 
Consulting Figs.~\ref{fig7}(a) and (b), one can notice that the definitions of $\tau$ are different between the Hahn echo and the dynamical decoupling.
In the former, $\tau$ is the time separation between the $\pi$/2 and $\pi$ pulses.
In the latter, $\tau$ is the separation between the adjacent $\pi$ pulses and the $(\pi/2)\mathrm{-}\pi$ separation becomes $\tau$/2 instead.
This is a widespread convention, and the benefit is that the matching condition for sensing an ac oscillation at frequency
$f_{\mathrm{ac}}$ becomes $f_{\mathrm{ac}} = (2\tau)^{-1}$ in both cases.

For dynamical decoupling, we use a family of XY sequences given by
XY4 = $\tau$/2--X--$\tau$--Y--$\tau$--X--$\tau$--Y--$\tau$/2, XY8 = XY4--YX4, XY16 = XY8--\={X}\={Y}8,
where X, Y, \={X}, and \={Y} denote the $\pi$ pulses around $x$, $y$, $-x$, and $-y$ axes of the rotating frame, respectively.~\cite{GBC90}
YX4 is XY4 with the positions of X's and Y's swapped (X $\leftrightarrow$ Y),
and \={X}\={Y}8 is XY8 with the rotation axes inverted (X $\to$ \={X}, Y $\to$ \={Y}) 
The XY$k$ ($k$ = 4, 8, 16) sequence repeated $N/k$ times is denoted as XY$k$-$N$, forming the \fbox{$N$} block in Fig.~\ref{fig7}(b).

The ground state of the NV center is spin-triplet, forming a qutrit system.
Although we usually pick up two states out of the three, treating it as a projected qubit system,
the qutrit nature of the NV ground state can have profound consequences on the relaxation processes.
The qubit spanned by the $m_S$ = 0 and 1 states or the $m_S$ = 0 and $-$1 states are magnetically allowed (SQ transition), 
and therefore the dominant relaxation source is also magnetic in origin.
The qubit spanned by the $m_S$ = $-$1 and 1 states on the other hand are magnetically forbidden (DQ transition),
but the Hamiltonian of the NV spin system has terms such as
$d_{\parallel} \Pi_{\parallel} S_z^2$ and $d_{\perp} \Pi_{\perp} (S_+^2 + S_-^2)/2$,
making the $m_S$ = $-$1 $\leftrightarrow$ 1 transition susceptible to electric field noise.
Therefore, a comparison of the SQ and DQ relaxation channels allows us to distinguish the prevailing noise source in the system.
Using the sequences in Fig.~\ref{fig7}(c), the normalized relaxation $S_{\mathrm{relax}}$ through the SQ and DQ relaxation channels are deduced, respectively, from
\begin{eqnarray}
S_{0,0} - S_{0,-1} &=& e^{ -\tau_{\mathrm{w}}/T_{1,\mathrm{SQ}} } \\
S_{-1,-1} - S_{-1,1} &=& e^{ -\tau_{\mathrm{w}}/T_{1,\mathrm{DQ}} }.
\end{eqnarray}
$T_{1,\mathrm{SQ}}$ is what we usually call the spin relaxation time $T_1$, but in practice includes contributions from both the SQ and DQ channels.
This is also the case for $T_{1,\mathrm{DQ}}$.
The relaxation times and the SQ and DQ relaxation rates ($\Omega$ and $\gamma$, respectively) are related by
$T_{1,\mathrm{SQ}} = (3 \Omega)^{-1}$ and $T_{1,\mathrm{DQ}} = (\Omega+2\gamma)^{-1}$.~\cite{MAJ17}

\section{Double electron--electron resonance\label{sec_deer}}
DEER is a technique that employs multifrequency to simultaneously drive two paramagnetic species.
Applied to the NV spin system, DEER provides a means to optically detect {\it dark} paramagnetic spins using NV centers as a probe spin.
Typically, the dark spins to be detected are P1 centers (substitutional N donors) or $g$-factor of $g_{\mathrm{e}}$ = 2 spins
due to divacancies (V$_2$) or surface defects/adatoms.
In these cases, there exist large frequency differences between the NV and paramagnetic spins, because of the large zero-field splitting of the ground state of the NV center.
At 23.4~mT applied below, $\omega_{0,-1}/2\pi$ is 2.215~GHz while $g_{\mathrm{e}}$ = 2 corresponds to 655~MHz.

The radiofrequency (RF) pulse tuned to the dark spin, applied amid the Hahn echo sequence on the NV spin [Fig.~\ref{fig7}(d)], leads to reduced coherence of the NV spin,
as it modifies a local magnetic environment of the NV spin dipolarly coupled with the dark spins.
The Hahn echo decay is modified as
\begin{equation}
\mathcal{C}_{\mathrm{DEER}} = \exp \left[ -\left( \frac{2\tau}{ T_{2,\mathrm{DEER}} }\right)^q \right].
\label{eq_deer}
\end{equation}
This mechanism is equivalent to the instantaneous diffusion, in which two paramagnetic spins share the same transition frequencies
and thus are flipped simultaneously by a single resonant pulse.
When the decay is single-exponential ($q$ = 1), the instantaneous diffusion time $T_{\mathrm{id}}$ is given by~\cite{SJ01,SKZ+17}
\begin{equation}
\frac{1}{ T_{\mathrm{id}} } = N_{\mathrm{d}} \frac{ \pi \mu_0 g_{\mathrm{e}}^2 \mu_{\mathrm{B}}^2 }{ 9\sqrt{3} \hbar } \sin^2 \left( \frac{\beta}{2} \right),
\label{eq_id}
\end{equation}
with $N_{\mathrm{d}}$ the density of the driven paramagnetic defects (assuming a uniform distribution),
$\mu_{\mathrm{B}}$ the Bohr magneton, and $\beta$ the flip angle.

The other operational mode of DEER is to continuously drive the paramagnetic spins throughout the Hahn echo of the NV spin [Fig.~\ref{fig7}(e)],
by which the dipolar interaction between the NV and paramagnetic spins is to be averaged out, and the coherence of the NV centers can be extended.~\cite{BZM+19}

\begin{figure}
\begin{center}
\includegraphics{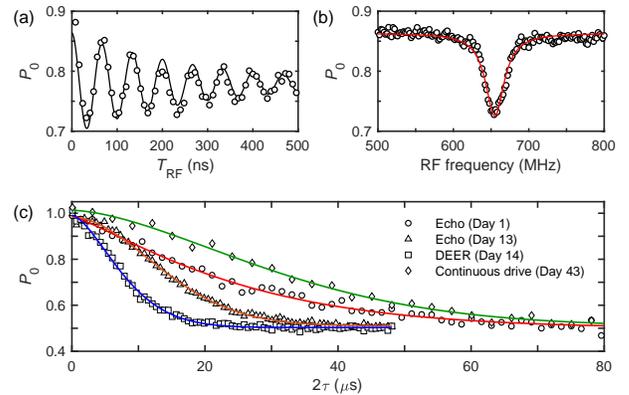}
\caption{(a) Rabi oscillation of paramagnetic defects driven at 655~MHz ($B_0$ = 23.4~mT).
(b) DEER spectrum.
(c) Coherence decays by Hahn echo ($\bigcirc$, $\triangle$), DEER ($\square$), and continuous driving ($\Diamond$).
Note that the respective data were taken different dates.
Counting the first Hahn echo ($\bigcirc$) as Day 1, other data were taken as specified in the figure legend.
\label{fig8}}
\end{center}
\end{figure}

DEER experiments were performed on various NV centers, but many, including the one measured in Sec.~\ref{sec_noise}, did not show clear spectra nor changes in $T_2$.
In fact, only one NV center, specified as the filled triangle in Fig.~\ref{fig2}(c), exhibited a clear spectrum.
DEER from this NV center is presented in Fig.~\ref{fig8}. 
The data also show an example of the instability of the spin coherence, often observed in shallow NV centers.
The initial Hahn echo showed $T_{2,\mathrm{echo}}$ = 26.6~$\mu$s ($\bigcirc$) 
and the depth was determined to be $d_{\mathrm{NV}}$ = 14.4~nm by the NMR measurement conducted immediately after.
When we returned to this NV center for pulsed DEER, $T_{2,\mathrm{echo}}$ was shortened to 16.9~$\mu$s ($\triangle$).
The pulsed DEER ($\square$) shows even shorter coherence, as expected.
The pulsed DEER data were fitted by Eqs.~(\ref{eq_t2}) and (\ref{eq_deer}),
using $T_{2,\mathrm{echo}}$ = 16.9~$\mu$s and $p$ = 1.14 obtained by the Hahn echo as fixed parameters.
We obtain $T_{2,\mathrm{DEER}}$ = 13.1~$\mu$s and $q$ = 1.91.
Naively, we estimate $N_{\mathrm{d}}$ to be on the order of 10$^{17}$~cm$^{-3}$, from Eq.~(\ref{eq_id}).
$N_{\mathrm{{d}}}$ of 10$^{17}$~cm$^{-3}$ corresponds to the average distance of about 20~nm, a value similar to $d_{\mathrm{NV}}$.

Continuous drive experiment was performed even later, but the coherence was more or less stable, with $T_{2,\mathrm{echo}}$ = 19.0~$\mu$s and $p$ = 1.76 (the data not shown).
This is extended to $T_{2,\mathrm{drive}}$ = 36.5~$\mu$s with $q$ = 1.64 ($\Diamond$).

It is not fully clear why we did not find DEER signals from many of the NV centers tested.
One clue is the short correlation time $\tau_{\mathrm{c}}$ associated with the broadband (white) noise observed in Sec.~\ref{sec_noise},
which indicates that the surface defect spins fluctuate in much faster timescale than the radiofrequency pulse can drive coherently.
It is then possible that this particular NV center with which DEER was observed might have coupled accidentally to defects not originating from the surface, such as V$_2$,
which should have a density much less than 10$^{17}$~cm$^{-3}$ globally but could have locally due to insufficient annealing.
This speculation is partly supported by the fact that $T_{2,\mathrm{drive}}$ of 36.5~$\mu$s under continuous driving is
still well within the observed $T_2$--$d_{\mathrm{NV}}$ trend in Fig.~\ref{fig2}(c), suggesting that $T_{2,\mathrm{drive}}$ is still limited by the noise source common to all the near-surface NV centers.

\bibliography{t2depth}

\begin{thebibliography}{52}%
\makeatletter
\providecommand \@ifxundefined [1]{%
 \@ifx{#1\undefined}
}%
\providecommand \@ifnum [1]{%
 \ifnum #1\expandafter \@firstoftwo
 \else \expandafter \@secondoftwo
 \fi
}%
\providecommand \@ifx [1]{%
 \ifx #1\expandafter \@firstoftwo
 \else \expandafter \@secondoftwo
 \fi
}%
\providecommand \natexlab [1]{#1}%
\providecommand \enquote  [1]{``#1''}%
\providecommand \bibnamefont  [1]{#1}%
\providecommand \bibfnamefont [1]{#1}%
\providecommand \citenamefont [1]{#1}%
\providecommand \href@noop [0]{\@secondoftwo}%
\providecommand \href [0]{\begingroup \@sanitize@url \@href}%
\providecommand \@href[1]{\@@startlink{#1}\@@href}%
\providecommand \@@href[1]{\endgroup#1\@@endlink}%
\providecommand \@sanitize@url [0]{\catcode `\\12\catcode `\$12\catcode
  `\&12\catcode `\#12\catcode `\^12\catcode `\_12\catcode `\%12\relax}%
\providecommand \@@startlink[1]{}%
\providecommand \@@endlink[0]{}%
\providecommand \url  [0]{\begingroup\@sanitize@url \@url }%
\providecommand \@url [1]{\endgroup\@href {#1}{\urlprefix }}%
\providecommand \urlprefix  [0]{URL }%
\providecommand \Eprint [0]{\href }%
\providecommand \doibase [0]{http://dx.doi.org/}%
\providecommand \selectlanguage [0]{\@gobble}%
\providecommand \bibinfo  [0]{\@secondoftwo}%
\providecommand \bibfield  [0]{\@secondoftwo}%
\providecommand \translation [1]{[#1]}%
\providecommand \BibitemOpen [0]{}%
\providecommand \bibitemStop [0]{}%
\providecommand \bibitemNoStop [0]{.\EOS\space}%
\providecommand \EOS [0]{\spacefactor3000\relax}%
\providecommand \BibitemShut  [1]{\csname bibitem#1\endcsname}%
\let\auto@bib@innerbib\@empty
\bibitem [{\citenamefont {Mamin}\ \emph {et~al.}(2013)\citenamefont {Mamin},
  \citenamefont {Kim}, \citenamefont {Sherwood}, \citenamefont {Rettner},
  \citenamefont {Ohno}, \citenamefont {Awschalom},\ and\ \citenamefont
  {Rugar}}]{MKS+13}%
  \BibitemOpen
  \bibfield  {author} {\bibinfo {author} {\bibfnamefont {H.~J.}\ \bibnamefont
  {Mamin}}, \bibinfo {author} {\bibfnamefont {M.}~\bibnamefont {Kim}}, \bibinfo
  {author} {\bibfnamefont {M.~H.}\ \bibnamefont {Sherwood}}, \bibinfo {author}
  {\bibfnamefont {C.~T.}\ \bibnamefont {Rettner}}, \bibinfo {author}
  {\bibfnamefont {K.}~\bibnamefont {Ohno}}, \bibinfo {author} {\bibfnamefont
  {D.~D.}\ \bibnamefont {Awschalom}}, \ and\ \bibinfo {author} {\bibfnamefont
  {D.}~\bibnamefont {Rugar}},\ }\bibfield  {title} {\enquote {\bibinfo {title}
  {Nanoscale {N}uclear {M}agnetic {R}esonance with a {N}itrogen-{V}acancy
  {S}pin {S}ensor},}\ }\href@noop {} {\bibfield  {journal} {\bibinfo  {journal}
  {Science}\ }\textbf {\bibinfo {volume} {339}},\ \bibinfo {pages} {557}
  (\bibinfo {year} {2013})}\BibitemShut {NoStop}%
\bibitem [{\citenamefont {Staudacher}\ \emph {et~al.}(2013)\citenamefont
  {Staudacher}, \citenamefont {Shi}, \citenamefont {Pezzagna}, \citenamefont
  {Meijer}, \citenamefont {Du}, \citenamefont {Meriles}, \citenamefont
  {Reinhard},\ and\ \citenamefont {Wrachtrup}}]{SSP+13}%
  \BibitemOpen
  \bibfield  {author} {\bibinfo {author} {\bibfnamefont {T.}~\bibnamefont
  {Staudacher}}, \bibinfo {author} {\bibfnamefont {F.}~\bibnamefont {Shi}},
  \bibinfo {author} {\bibfnamefont {S.}~\bibnamefont {Pezzagna}}, \bibinfo
  {author} {\bibfnamefont {J.}~\bibnamefont {Meijer}}, \bibinfo {author}
  {\bibfnamefont {J.}~\bibnamefont {Du}}, \bibinfo {author} {\bibfnamefont
  {C.~A.}\ \bibnamefont {Meriles}}, \bibinfo {author} {\bibfnamefont
  {F.}~\bibnamefont {Reinhard}}, \ and\ \bibinfo {author} {\bibfnamefont
  {J.}~\bibnamefont {Wrachtrup}},\ }\bibfield  {title} {\enquote {\bibinfo
  {title} {Nuclear {M}agnetic {R}esonance {S}pectroscopy on a
  (5-{N}anometer)$^3$ {S}ample {V}olume},}\ }\href@noop {} {\bibfield
  {journal} {\bibinfo  {journal} {Science}\ }\textbf {\bibinfo {volume}
  {339}},\ \bibinfo {pages} {561} (\bibinfo {year} {2013})}\BibitemShut
  {NoStop}%
\bibitem [{\citenamefont {M{\"u}ller}\ \emph {et~al.}(2014)\citenamefont
  {M{\"u}ller}, \citenamefont {Kong}, \citenamefont {Cai}, \citenamefont
  {Melentijevi{\'c}}, \citenamefont {Stacey}, \citenamefont {Markham},
  \citenamefont {Twitchen}, \citenamefont {Isoya}, \citenamefont {Pezzagna},
  \citenamefont {Meijer}, \citenamefont {Du}, \citenamefont {Plenio},
  \citenamefont {Naydenov}, \citenamefont {McGuinness},\ and\ \citenamefont
  {Jelezko}}]{MKC+14}%
  \BibitemOpen
  \bibfield  {author} {\bibinfo {author} {\bibfnamefont {C.}~\bibnamefont
  {M{\"u}ller}}, \bibinfo {author} {\bibfnamefont {X.}~\bibnamefont {Kong}},
  \bibinfo {author} {\bibfnamefont {J.-M.}\ \bibnamefont {Cai}}, \bibinfo
  {author} {\bibfnamefont {K.}~\bibnamefont {Melentijevi{\'c}}}, \bibinfo
  {author} {\bibfnamefont {A.}~\bibnamefont {Stacey}}, \bibinfo {author}
  {\bibfnamefont {M.}~\bibnamefont {Markham}}, \bibinfo {author} {\bibfnamefont
  {D.}~\bibnamefont {Twitchen}}, \bibinfo {author} {\bibfnamefont
  {J.}~\bibnamefont {Isoya}}, \bibinfo {author} {\bibfnamefont
  {S.}~\bibnamefont {Pezzagna}}, \bibinfo {author} {\bibfnamefont
  {J.}~\bibnamefont {Meijer}}, \bibinfo {author} {\bibfnamefont {J.~F.}\
  \bibnamefont {Du}}, \bibinfo {author} {\bibfnamefont {M.~B.}\ \bibnamefont
  {Plenio}}, \bibinfo {author} {\bibfnamefont {B.}~\bibnamefont {Naydenov}},
  \bibinfo {author} {\bibfnamefont {L.~P.}\ \bibnamefont {McGuinness}}, \ and\
  \bibinfo {author} {\bibfnamefont {F.}~\bibnamefont {Jelezko}},\ }\bibfield
  {title} {\enquote {\bibinfo {title} {Nuclear magnetic resonance spectroscopy
  with single spin sensitivity},}\ }\href@noop {} {\bibfield  {journal}
  {\bibinfo  {journal} {Nat.\ Commun.}\ }\textbf {\bibinfo {volume} {5}},\
  \bibinfo {pages} {4703} (\bibinfo {year} {2014})}\BibitemShut {NoStop}%
\bibitem [{\citenamefont {H{\"a}berle}\ \emph {et~al.}(2015)\citenamefont
  {H{\"a}berle}, \citenamefont {Schmid-Lorch}, \citenamefont {Reinhard},\ and\
  \citenamefont {Wrachtrup}}]{HSR+15}%
  \BibitemOpen
  \bibfield  {author} {\bibinfo {author} {\bibfnamefont {T.}~\bibnamefont
  {H{\"a}berle}}, \bibinfo {author} {\bibfnamefont {D.}~\bibnamefont
  {Schmid-Lorch}}, \bibinfo {author} {\bibfnamefont {F.}~\bibnamefont
  {Reinhard}}, \ and\ \bibinfo {author} {\bibfnamefont {J.}~\bibnamefont
  {Wrachtrup}},\ }\bibfield  {title} {\enquote {\bibinfo {title} {Nanoscale
  nuclear magnetic imaging with chemical contrast},}\ }\href@noop {} {\bibfield
   {journal} {\bibinfo  {journal} {Nat.\ Nanotechnol.}\ }\textbf {\bibinfo
  {volume} {10}},\ \bibinfo {pages} {125} (\bibinfo {year} {2015})}\BibitemShut
  {NoStop}%
\bibitem [{\citenamefont {DeVience}\ \emph {et~al.}(2015)\citenamefont
  {DeVience}, \citenamefont {Pham}, \citenamefont {Lovchinsky}, \citenamefont
  {Sushkov}, \citenamefont {Bar-Gill}, \citenamefont {Belthangady},
  \citenamefont {Casola}, \citenamefont {Corbett}, \citenamefont {Zhang},
  \citenamefont {Lukin}, \citenamefont {Park}, \citenamefont {Yacoby},\ and\
  \citenamefont {Walsworth}}]{DPL+15}%
  \BibitemOpen
  \bibfield  {author} {\bibinfo {author} {\bibfnamefont {S.~J.}\ \bibnamefont
  {DeVience}}, \bibinfo {author} {\bibfnamefont {L.~M.}\ \bibnamefont {Pham}},
  \bibinfo {author} {\bibfnamefont {I.}~\bibnamefont {Lovchinsky}}, \bibinfo
  {author} {\bibfnamefont {A.~O.}\ \bibnamefont {Sushkov}}, \bibinfo {author}
  {\bibfnamefont {N.}~\bibnamefont {Bar-Gill}}, \bibinfo {author}
  {\bibfnamefont {C.}~\bibnamefont {Belthangady}}, \bibinfo {author}
  {\bibfnamefont {F.}~\bibnamefont {Casola}}, \bibinfo {author} {\bibfnamefont
  {M.}~\bibnamefont {Corbett}}, \bibinfo {author} {\bibfnamefont
  {H.}~\bibnamefont {Zhang}}, \bibinfo {author} {\bibfnamefont
  {M.}~\bibnamefont {Lukin}}, \bibinfo {author} {\bibfnamefont
  {H.}~\bibnamefont {Park}}, \bibinfo {author} {\bibfnamefont {A.}~\bibnamefont
  {Yacoby}}, \ and\ \bibinfo {author} {\bibfnamefont {R.~L.}\ \bibnamefont
  {Walsworth}},\ }\bibfield  {title} {\enquote {\bibinfo {title} {Nanoscale
  {NMR} spectroscopy and imaging of multiple nulcear species},}\ }\href@noop {}
  {\bibfield  {journal} {\bibinfo  {journal} {Nat.\ Nanotechnol.}\ }\textbf
  {\bibinfo {volume} {10}},\ \bibinfo {pages} {129} (\bibinfo {year}
  {2015})}\BibitemShut {NoStop}%
\bibitem [{\citenamefont {Lovchinsky}\ \emph {et~al.}(2016)\citenamefont
  {Lovchinsky}, \citenamefont {Sushkov}, \citenamefont {Urbach}, \citenamefont
  {de~Leon}, \citenamefont {Choi}, \citenamefont {De~Greve}, \citenamefont
  {Evans}, \citenamefont {Gertner}, \citenamefont {Bersin}, \citenamefont
  {M{\"u}ller}, \citenamefont {McGuinness}, \citenamefont {Jelezko},
  \citenamefont {Walsworth}, \citenamefont {Park},\ and\ \citenamefont
  {Lukin}}]{LSU+16}%
  \BibitemOpen
  \bibfield  {author} {\bibinfo {author} {\bibfnamefont {I.}~\bibnamefont
  {Lovchinsky}}, \bibinfo {author} {\bibfnamefont {A.~O.}\ \bibnamefont
  {Sushkov}}, \bibinfo {author} {\bibfnamefont {E.}~\bibnamefont {Urbach}},
  \bibinfo {author} {\bibfnamefont {N.~P.}\ \bibnamefont {de~Leon}}, \bibinfo
  {author} {\bibfnamefont {S.}~\bibnamefont {Choi}}, \bibinfo {author}
  {\bibfnamefont {K.}~\bibnamefont {De~Greve}}, \bibinfo {author}
  {\bibfnamefont {R.}~\bibnamefont {Evans}}, \bibinfo {author} {\bibfnamefont
  {R.}~\bibnamefont {Gertner}}, \bibinfo {author} {\bibfnamefont
  {E.}~\bibnamefont {Bersin}}, \bibinfo {author} {\bibfnamefont
  {C.}~\bibnamefont {M{\"u}ller}}, \bibinfo {author} {\bibfnamefont
  {L.}~\bibnamefont {McGuinness}}, \bibinfo {author} {\bibfnamefont
  {F.}~\bibnamefont {Jelezko}}, \bibinfo {author} {\bibfnamefont {R.~L.}\
  \bibnamefont {Walsworth}}, \bibinfo {author} {\bibfnamefont {H.}~\bibnamefont
  {Park}}, \ and\ \bibinfo {author} {\bibfnamefont {M.~D.}\ \bibnamefont
  {Lukin}},\ }\bibfield  {title} {\enquote {\bibinfo {title} {Nuclear magnetic
  resonance detection and spectroscopy of single proteins using quantum
  logic},}\ }\href@noop {} {\bibfield  {journal} {\bibinfo  {journal}
  {Science}\ }\textbf {\bibinfo {volume} {351}},\ \bibinfo {pages} {836}
  (\bibinfo {year} {2016})}\BibitemShut {NoStop}%
\bibitem [{\citenamefont {Aslam}\ \emph {et~al.}(2017)\citenamefont {Aslam},
  \citenamefont {Pfender}, \citenamefont {Neumann}, \citenamefont {Reuter},
  \citenamefont {Zappe}, \citenamefont {de~Oliveira}, \citenamefont
  {Denisenko}, \citenamefont {Sumiya}, \citenamefont {Onoda}, \citenamefont
  {Isoya},\ and\ \citenamefont {Wrachtrup}}]{APN+17}%
  \BibitemOpen
  \bibfield  {author} {\bibinfo {author} {\bibfnamefont {N.}~\bibnamefont
  {Aslam}}, \bibinfo {author} {\bibfnamefont {M.}~\bibnamefont {Pfender}},
  \bibinfo {author} {\bibfnamefont {P.}~\bibnamefont {Neumann}}, \bibinfo
  {author} {\bibfnamefont {R.}~\bibnamefont {Reuter}}, \bibinfo {author}
  {\bibfnamefont {A.}~\bibnamefont {Zappe}}, \bibinfo {author} {\bibfnamefont
  {F.~F.}\ \bibnamefont {de~Oliveira}}, \bibinfo {author} {\bibfnamefont
  {A.}~\bibnamefont {Denisenko}}, \bibinfo {author} {\bibfnamefont
  {H.}~\bibnamefont {Sumiya}}, \bibinfo {author} {\bibfnamefont
  {S.}~\bibnamefont {Onoda}}, \bibinfo {author} {\bibfnamefont
  {J.}~\bibnamefont {Isoya}}, \ and\ \bibinfo {author} {\bibfnamefont
  {J.}~\bibnamefont {Wrachtrup}},\ }\bibfield  {title} {\enquote {\bibinfo
  {title} {Nanoscale nuclear magnetic resonance with chemical resolution},}\
  }\href@noop {} {\bibfield  {journal} {\bibinfo  {journal} {Science}\ }\textbf
  {\bibinfo {volume} {357}},\ \bibinfo {pages} {67} (\bibinfo {year}
  {2017})}\BibitemShut {NoStop}%
\bibitem [{\citenamefont {Tame}\ \emph {et~al.}(2013)\citenamefont {Tame},
  \citenamefont {McEnery}, \citenamefont {{\"O}zdemir}, \citenamefont {Lee},
  \citenamefont {Maier},\ and\ \citenamefont {Kim}}]{TMO+13}%
  \BibitemOpen
  \bibfield  {author} {\bibinfo {author} {\bibfnamefont {M.~S.}\ \bibnamefont
  {Tame}}, \bibinfo {author} {\bibfnamefont {K.~R.}\ \bibnamefont {McEnery}},
  \bibinfo {author} {\bibfnamefont {K.}~\bibnamefont {{\"O}zdemir}}, \bibinfo
  {author} {\bibfnamefont {J.}~\bibnamefont {Lee}}, \bibinfo {author}
  {\bibfnamefont {S.~A.}\ \bibnamefont {Maier}}, \ and\ \bibinfo {author}
  {\bibfnamefont {M.~S.}\ \bibnamefont {Kim}},\ }\bibfield  {title} {\enquote
  {\bibinfo {title} {Quantum plasmonics},}\ }\href@noop {} {\bibfield
  {journal} {\bibinfo  {journal} {Nat.\ Phys.}\ }\textbf {\bibinfo {volume}
  {9}},\ \bibinfo {pages} {329} (\bibinfo {year} {2013})}\BibitemShut {NoStop}%
\bibitem [{\citenamefont {Schr{\"o}der}\ \emph {et~al.}(2016)\citenamefont
  {Schr{\"o}der}, \citenamefont {Mouradian}, \citenamefont {Zheng},
  \citenamefont {Trusheim}, \citenamefont {Walsh}, \citenamefont {Chen},
  \citenamefont {Li}, \citenamefont {Bayn},\ and\ \citenamefont
  {Englund}}]{SMZ+16}%
  \BibitemOpen
  \bibfield  {author} {\bibinfo {author} {\bibfnamefont {T.}~\bibnamefont
  {Schr{\"o}der}}, \bibinfo {author} {\bibfnamefont {S.~L.}\ \bibnamefont
  {Mouradian}}, \bibinfo {author} {\bibfnamefont {J.}~\bibnamefont {Zheng}},
  \bibinfo {author} {\bibfnamefont {M.~E.}\ \bibnamefont {Trusheim}}, \bibinfo
  {author} {\bibfnamefont {M.}~\bibnamefont {Walsh}}, \bibinfo {author}
  {\bibfnamefont {E.~H.}\ \bibnamefont {Chen}}, \bibinfo {author}
  {\bibfnamefont {L.}~\bibnamefont {Li}}, \bibinfo {author} {\bibfnamefont
  {I.}~\bibnamefont {Bayn}}, \ and\ \bibinfo {author} {\bibfnamefont
  {D.}~\bibnamefont {Englund}},\ }\bibfield  {title} {\enquote {\bibinfo
  {title} {Quantum nanophotonics in diamond},}\ }\href@noop {} {\bibfield
  {journal} {\bibinfo  {journal} {J.\ Opt.\ Soc.\ Am.\ B}\ }\textbf {\bibinfo
  {volume} {33}},\ \bibinfo {pages} {B65} (\bibinfo {year} {2016})}\BibitemShut
  {NoStop}%
\bibitem [{\citenamefont {Atat{\"u}re}\ \emph {et~al.}(2018)\citenamefont
  {Atat{\"u}re}, \citenamefont {Englund}, \citenamefont {Vamivakas},
  \citenamefont {Lee},\ and\ \citenamefont {Wrachtrup}}]{AEV+18}%
  \BibitemOpen
  \bibfield  {author} {\bibinfo {author} {\bibfnamefont {M.}~\bibnamefont
  {Atat{\"u}re}}, \bibinfo {author} {\bibfnamefont {D.}~\bibnamefont
  {Englund}}, \bibinfo {author} {\bibfnamefont {N.}~\bibnamefont {Vamivakas}},
  \bibinfo {author} {\bibfnamefont {S.-Y.}\ \bibnamefont {Lee}}, \ and\
  \bibinfo {author} {\bibfnamefont {J.}~\bibnamefont {Wrachtrup}},\ }\bibfield
  {title} {\enquote {\bibinfo {title} {Material platforms for spin-based
  photonic quantum technologies},}\ }\href@noop {} {\bibfield  {journal}
  {\bibinfo  {journal} {Nat.\ Rev.\ Mater.}\ }\textbf {\bibinfo {volume} {3}},\
  \bibinfo {pages} {38} (\bibinfo {year} {2018})}\BibitemShut {NoStop}%
\bibitem [{\citenamefont {Awschalom}\ \emph {et~al.}(2018)\citenamefont
  {Awschalom}, \citenamefont {Hanson}, \citenamefont {Wrachtrup},\ and\
  \citenamefont {Zhou}}]{AHWZ18}%
  \BibitemOpen
  \bibfield  {author} {\bibinfo {author} {\bibfnamefont {D.~D.}\ \bibnamefont
  {Awschalom}}, \bibinfo {author} {\bibfnamefont {R.}~\bibnamefont {Hanson}},
  \bibinfo {author} {\bibfnamefont {J.}~\bibnamefont {Wrachtrup}}, \ and\
  \bibinfo {author} {\bibfnamefont {B.~B.}\ \bibnamefont {Zhou}},\ }\bibfield
  {title} {\enquote {\bibinfo {title} {Quantum technologies with optically
  interfaced solid-state spins},}\ }\href@noop {} {\bibfield  {journal}
  {\bibinfo  {journal} {Nat.\ Photon.}\ }\textbf {\bibinfo {volume} {12}},\
  \bibinfo {pages} {516} (\bibinfo {year} {2018})}\BibitemShut {NoStop}%
\bibitem [{\citenamefont {Wehner}, \citenamefont {Elkouss},\ and\ \citenamefont
  {Hanson}(2018)}]{WEH18}%
  \BibitemOpen
  \bibfield  {author} {\bibinfo {author} {\bibfnamefont {S.}~\bibnamefont
  {Wehner}}, \bibinfo {author} {\bibfnamefont {D.}~\bibnamefont {Elkouss}}, \
  and\ \bibinfo {author} {\bibfnamefont {R.}~\bibnamefont {Hanson}},\
  }\bibfield  {title} {\enquote {\bibinfo {title} {Quantum internet: {A} vision
  for the road ahead},}\ }\href@noop {} {\bibfield  {journal} {\bibinfo
  {journal} {Science}\ }\textbf {\bibinfo {volume} {362}},\ \bibinfo {pages}
  {eaam9288} (\bibinfo {year} {2018})}\BibitemShut {NoStop}%
\bibitem [{\citenamefont {Bradac}\ \emph {et~al.}(2019)\citenamefont {Bradac},
  \citenamefont {Gao}, \citenamefont {Forneris}, \citenamefont {Trusheim},\
  and\ \citenamefont {Aharonovich}}]{BGF+19}%
  \BibitemOpen
  \bibfield  {author} {\bibinfo {author} {\bibfnamefont {C.}~\bibnamefont
  {Bradac}}, \bibinfo {author} {\bibfnamefont {W.}~\bibnamefont {Gao}},
  \bibinfo {author} {\bibfnamefont {J.}~\bibnamefont {Forneris}}, \bibinfo
  {author} {\bibfnamefont {M.~E.}\ \bibnamefont {Trusheim}}, \ and\ \bibinfo
  {author} {\bibfnamefont {I.}~\bibnamefont {Aharonovich}},\ }\bibfield
  {title} {\enquote {\bibinfo {title} {Quantum nanophotonics with group {IV}
  defects in diamond},}\ }\href@noop {} {\bibfield  {journal} {\bibinfo
  {journal} {Nat.\ Commun.}\ }\textbf {\bibinfo {volume} {10}},\ \bibinfo
  {pages} {5625} (\bibinfo {year} {2019})}\BibitemShut {NoStop}%
\bibitem [{\citenamefont {Pezzagna}\ \emph {et~al.}(2010)\citenamefont
  {Pezzagna}, \citenamefont {Naydenov}, \citenamefont {Jelezko}, \citenamefont
  {Wrachtrup},\ and\ \citenamefont {Meijer}}]{PNJ+10}%
  \BibitemOpen
  \bibfield  {author} {\bibinfo {author} {\bibfnamefont {S.}~\bibnamefont
  {Pezzagna}}, \bibinfo {author} {\bibfnamefont {B.}~\bibnamefont {Naydenov}},
  \bibinfo {author} {\bibfnamefont {F.}~\bibnamefont {Jelezko}}, \bibinfo
  {author} {\bibfnamefont {J.}~\bibnamefont {Wrachtrup}}, \ and\ \bibinfo
  {author} {\bibfnamefont {J.}~\bibnamefont {Meijer}},\ }\bibfield  {title}
  {\enquote {\bibinfo {title} {Creation efficiency of nitrogen-vacancy centres
  in diamond},}\ }\href@noop {} {\bibfield  {journal} {\bibinfo  {journal} {New
  J.\ Phys.}\ }\textbf {\bibinfo {volume} {12}},\ \bibinfo {pages} {065017}
  (\bibinfo {year} {2010})}\BibitemShut {NoStop}%
\bibitem [{\citenamefont {Ofori-Okai}\ \emph {et~al.}(2012)\citenamefont
  {Ofori-Okai}, \citenamefont {Pezzagna}, \citenamefont {Chang}, \citenamefont
  {Loretz}, \citenamefont {Schirhagl}, \citenamefont {Tao}, \citenamefont
  {Moores}, \citenamefont {Groot-Berning}, \citenamefont {Meijer},\ and\
  \citenamefont {Degen}}]{OPC+12}%
  \BibitemOpen
  \bibfield  {author} {\bibinfo {author} {\bibfnamefont {B.~K.}\ \bibnamefont
  {Ofori-Okai}}, \bibinfo {author} {\bibfnamefont {S.}~\bibnamefont
  {Pezzagna}}, \bibinfo {author} {\bibfnamefont {K.}~\bibnamefont {Chang}},
  \bibinfo {author} {\bibfnamefont {M.}~\bibnamefont {Loretz}}, \bibinfo
  {author} {\bibfnamefont {R.}~\bibnamefont {Schirhagl}}, \bibinfo {author}
  {\bibfnamefont {Y.}~\bibnamefont {Tao}}, \bibinfo {author} {\bibfnamefont
  {B.~A.}\ \bibnamefont {Moores}}, \bibinfo {author} {\bibfnamefont
  {K.}~\bibnamefont {Groot-Berning}}, \bibinfo {author} {\bibfnamefont
  {J.}~\bibnamefont {Meijer}}, \ and\ \bibinfo {author} {\bibfnamefont {C.~L.}\
  \bibnamefont {Degen}},\ }\bibfield  {title} {\enquote {\bibinfo {title} {Spin
  properties of very shallow nitrogen vacancy defects in diamond},}\
  }\href@noop {} {\bibfield  {journal} {\bibinfo  {journal} {Phys.\ Rev.\ B}\
  }\textbf {\bibinfo {volume} {86}},\ \bibinfo {pages} {081406} (\bibinfo
  {year} {2012})}\BibitemShut {NoStop}%
\bibitem [{\citenamefont {de~Oliveira}\ \emph {et~al.}(2017)\citenamefont
  {de~Oliveira}, \citenamefont {Antonov}, \citenamefont {Wang}, \citenamefont
  {Neumann}, \citenamefont {Momenzadeh}, \citenamefont {H{\"a}u{\ss}ermann},
  \citenamefont {Pasquarelli}, \citenamefont {Denisenko},\ and\ \citenamefont
  {Wrachtrup}}]{dOAW+17}%
  \BibitemOpen
  \bibfield  {author} {\bibinfo {author} {\bibfnamefont {F.~F.}\ \bibnamefont
  {de~Oliveira}}, \bibinfo {author} {\bibfnamefont {D.}~\bibnamefont
  {Antonov}}, \bibinfo {author} {\bibfnamefont {Y.}~\bibnamefont {Wang}},
  \bibinfo {author} {\bibfnamefont {P.}~\bibnamefont {Neumann}}, \bibinfo
  {author} {\bibfnamefont {S.~A.}\ \bibnamefont {Momenzadeh}}, \bibinfo
  {author} {\bibfnamefont {T.}~\bibnamefont {H{\"a}u{\ss}ermann}}, \bibinfo
  {author} {\bibfnamefont {A.}~\bibnamefont {Pasquarelli}}, \bibinfo {author}
  {\bibfnamefont {A.}~\bibnamefont {Denisenko}}, \ and\ \bibinfo {author}
  {\bibfnamefont {J.}~\bibnamefont {Wrachtrup}},\ }\bibfield  {title} {\enquote
  {\bibinfo {title} {Tailoring spin defects in diamond by lattice charging},}\
  }\href@noop {} {\bibfield  {journal} {\bibinfo  {journal} {Nat.\ Commun.}\
  }\textbf {\bibinfo {volume} {8}},\ \bibinfo {pages} {15409} (\bibinfo {year}
  {2017})}\BibitemShut {NoStop}%
\bibitem [{\citenamefont {Fukuda}\ \emph {et~al.}(2018)\citenamefont {Fukuda},
  \citenamefont {Balasubramanian}, \citenamefont {Higashimata}, \citenamefont
  {Koike}, \citenamefont {Okada}, \citenamefont {Kagami}, \citenamefont
  {Teraji}, \citenamefont {Onoda}, \citenamefont {Haruyama}, \citenamefont
  {Yamada}, \citenamefont {Inaba}, \citenamefont {Yamano}, \citenamefont
  {St{\"u}rner}, \citenamefont {Schmitt}, \citenamefont {McGuinness},
  \citenamefont {Jelezko}, \citenamefont {Ohshima}, \citenamefont {Shinada},
  \citenamefont {Kawarada}, \citenamefont {Kada}, \citenamefont {Hanaizumi},
  \citenamefont {Tanii},\ and\ \citenamefont {Isoya}}]{FBH+18}%
  \BibitemOpen
  \bibfield  {author} {\bibinfo {author} {\bibfnamefont {R.}~\bibnamefont
  {Fukuda}}, \bibinfo {author} {\bibfnamefont {P.}~\bibnamefont
  {Balasubramanian}}, \bibinfo {author} {\bibfnamefont {I.}~\bibnamefont
  {Higashimata}}, \bibinfo {author} {\bibfnamefont {G.}~\bibnamefont {Koike}},
  \bibinfo {author} {\bibfnamefont {T.}~\bibnamefont {Okada}}, \bibinfo
  {author} {\bibfnamefont {R.}~\bibnamefont {Kagami}}, \bibinfo {author}
  {\bibfnamefont {T.}~\bibnamefont {Teraji}}, \bibinfo {author} {\bibfnamefont
  {S.}~\bibnamefont {Onoda}}, \bibinfo {author} {\bibfnamefont
  {M.}~\bibnamefont {Haruyama}}, \bibinfo {author} {\bibfnamefont
  {K.}~\bibnamefont {Yamada}}, \bibinfo {author} {\bibfnamefont
  {M.}~\bibnamefont {Inaba}}, \bibinfo {author} {\bibfnamefont
  {H.}~\bibnamefont {Yamano}}, \bibinfo {author} {\bibfnamefont {F.~M.}\
  \bibnamefont {St{\"u}rner}}, \bibinfo {author} {\bibfnamefont
  {S.}~\bibnamefont {Schmitt}}, \bibinfo {author} {\bibfnamefont {L.~P.}\
  \bibnamefont {McGuinness}}, \bibinfo {author} {\bibfnamefont
  {F.}~\bibnamefont {Jelezko}}, \bibinfo {author} {\bibfnamefont
  {T.}~\bibnamefont {Ohshima}}, \bibinfo {author} {\bibfnamefont
  {T.}~\bibnamefont {Shinada}}, \bibinfo {author} {\bibfnamefont
  {H.}~\bibnamefont {Kawarada}}, \bibinfo {author} {\bibfnamefont
  {W.}~\bibnamefont {Kada}}, \bibinfo {author} {\bibfnamefont {O.}~\bibnamefont
  {Hanaizumi}}, \bibinfo {author} {\bibfnamefont {T.}~\bibnamefont {Tanii}}, \
  and\ \bibinfo {author} {\bibfnamefont {J.}~\bibnamefont {Isoya}},\ }\bibfield
   {title} {\enquote {\bibinfo {title} {Lithographically engineered shallow
  nitrogen-vacancy centers in diamond for external nuclear spin sensing},}\
  }\href@noop {} {\bibfield  {journal} {\bibinfo  {journal} {New J.\ Phys.}\
  }\textbf {\bibinfo {volume} {20}},\ \bibinfo {pages} {083029} (\bibinfo
  {year} {2018})}\BibitemShut {NoStop}%
\bibitem [{\citenamefont {Sangtawesin}\ \emph {et~al.}(2019)\citenamefont
  {Sangtawesin}, \citenamefont {Dwyer}, \citenamefont {Srinivasan},
  \citenamefont {Allred}, \citenamefont {Rodgers}, \citenamefont {De~Greve},
  \citenamefont {Stacey}, \citenamefont {Dontschuk}, \citenamefont {O'Donnell},
  \citenamefont {Hu}, \citenamefont {Evans}, \citenamefont {Jaye},
  \citenamefont {Fischer}, \citenamefont {Markham}, \citenamefont {Twitchen},
  \citenamefont {Park}, \citenamefont {Lukin},\ and\ \citenamefont
  {de~Leon}}]{SDS+19}%
  \BibitemOpen
  \bibfield  {author} {\bibinfo {author} {\bibfnamefont {S.}~\bibnamefont
  {Sangtawesin}}, \bibinfo {author} {\bibfnamefont {B.~L.}\ \bibnamefont
  {Dwyer}}, \bibinfo {author} {\bibfnamefont {S.}~\bibnamefont {Srinivasan}},
  \bibinfo {author} {\bibfnamefont {J.~J.}\ \bibnamefont {Allred}}, \bibinfo
  {author} {\bibfnamefont {L.~V.~H.}\ \bibnamefont {Rodgers}}, \bibinfo
  {author} {\bibfnamefont {K.}~\bibnamefont {De~Greve}}, \bibinfo {author}
  {\bibfnamefont {A.}~\bibnamefont {Stacey}}, \bibinfo {author} {\bibfnamefont
  {N.}~\bibnamefont {Dontschuk}}, \bibinfo {author} {\bibfnamefont {K.~M.}\
  \bibnamefont {O'Donnell}}, \bibinfo {author} {\bibfnamefont {D.}~\bibnamefont
  {Hu}}, \bibinfo {author} {\bibfnamefont {D.~A.}\ \bibnamefont {Evans}},
  \bibinfo {author} {\bibfnamefont {C.}~\bibnamefont {Jaye}}, \bibinfo {author}
  {\bibfnamefont {D.~A.}\ \bibnamefont {Fischer}}, \bibinfo {author}
  {\bibfnamefont {M.~L.}\ \bibnamefont {Markham}}, \bibinfo {author}
  {\bibfnamefont {D.~J.}\ \bibnamefont {Twitchen}}, \bibinfo {author}
  {\bibfnamefont {H.}~\bibnamefont {Park}}, \bibinfo {author} {\bibfnamefont
  {M.~D.}\ \bibnamefont {Lukin}}, \ and\ \bibinfo {author} {\bibfnamefont
  {N.~P.}\ \bibnamefont {de~Leon}},\ }\bibfield  {title} {\enquote {\bibinfo
  {title} {Origins of {D}iamond {S}urface {N}oise {P}robed by {C}orrelating
  {S}ingle-{S}pin {M}easurements with {S}urface {S}pectroscopyy},}\ }\href@noop
  {} {\bibfield  {journal} {\bibinfo  {journal} {Phys.\ Rev.\ X}\ }\textbf
  {\bibinfo {volume} {9}},\ \bibinfo {pages} {031052} (\bibinfo {year}
  {2019})}\BibitemShut {NoStop}%
\bibitem [{\citenamefont {Ohno}\ \emph {et~al.}(2012)\citenamefont {Ohno},
  \citenamefont {Heremans}, \citenamefont {Bassett}, \citenamefont {Myers},
  \citenamefont {Toyli}, \citenamefont {Bleszynski~Jayich}, \citenamefont
  {Palmstr{\/o}m},\ and\ \citenamefont {Awschalom}}]{OHB+12}%
  \BibitemOpen
  \bibfield  {author} {\bibinfo {author} {\bibfnamefont {K.}~\bibnamefont
  {Ohno}}, \bibinfo {author} {\bibfnamefont {F.~J.}\ \bibnamefont {Heremans}},
  \bibinfo {author} {\bibfnamefont {L.~C.}\ \bibnamefont {Bassett}}, \bibinfo
  {author} {\bibfnamefont {B.~A.}\ \bibnamefont {Myers}}, \bibinfo {author}
  {\bibfnamefont {D.~M.}\ \bibnamefont {Toyli}}, \bibinfo {author}
  {\bibfnamefont {A.~C.}\ \bibnamefont {Bleszynski~Jayich}}, \bibinfo {author}
  {\bibfnamefont {C.~J.}\ \bibnamefont {Palmstr{\/o}m}}, \ and\ \bibinfo
  {author} {\bibfnamefont {D.~D.}\ \bibnamefont {Awschalom}},\ }\bibfield
  {title} {\enquote {\bibinfo {title} {Engineering shallow spins in diamond
  with nitrogen delta-doping},}\ }\href@noop {} {\bibfield  {journal} {\bibinfo
   {journal} {Appl.\ Phys.\ Lett.}\ }\textbf {\bibinfo {volume} {101}},\
  \bibinfo {pages} {082413} (\bibinfo {year} {2012})}\BibitemShut {NoStop}%
\bibitem [{\citenamefont {Ohashi}\ \emph {et~al.}(2013)\citenamefont {Ohashi},
  \citenamefont {Rosskopf}, \citenamefont {Watanabe}, \citenamefont {Loretz},
  \citenamefont {Tao}, \citenamefont {Hauert}, \citenamefont {Tomizawa},
  \citenamefont {Ishikawa}, \citenamefont {Ishi-Hayase}, \citenamefont
  {Shikata}, \citenamefont {Degen},\ and\ \citenamefont {Itoh}}]{ORW+13}%
  \BibitemOpen
  \bibfield  {author} {\bibinfo {author} {\bibfnamefont {K.}~\bibnamefont
  {Ohashi}}, \bibinfo {author} {\bibfnamefont {T.}~\bibnamefont {Rosskopf}},
  \bibinfo {author} {\bibfnamefont {H.}~\bibnamefont {Watanabe}}, \bibinfo
  {author} {\bibfnamefont {M.}~\bibnamefont {Loretz}}, \bibinfo {author}
  {\bibfnamefont {Y.}~\bibnamefont {Tao}}, \bibinfo {author} {\bibfnamefont
  {R.}~\bibnamefont {Hauert}}, \bibinfo {author} {\bibfnamefont
  {S.}~\bibnamefont {Tomizawa}}, \bibinfo {author} {\bibfnamefont
  {T.}~\bibnamefont {Ishikawa}}, \bibinfo {author} {\bibfnamefont
  {J.}~\bibnamefont {Ishi-Hayase}}, \bibinfo {author} {\bibfnamefont
  {S.}~\bibnamefont {Shikata}}, \bibinfo {author} {\bibfnamefont {C.~L.}\
  \bibnamefont {Degen}}, \ and\ \bibinfo {author} {\bibfnamefont {K.~M.}\
  \bibnamefont {Itoh}},\ }\bibfield  {title} {\enquote {\bibinfo {title}
  {Negatively {C}harged {N}itrogen-{V}acancy {C}enters in a 5 nm {T}hin
  $^{12}${C} {D}iamond {F}ilm},}\ }\href@noop {} {\bibfield  {journal}
  {\bibinfo  {journal} {Nano Lett.}\ }\textbf {\bibinfo {volume} {13}},\
  \bibinfo {pages} {4733} (\bibinfo {year} {2013})}\BibitemShut {NoStop}%
\bibitem [{\citenamefont {Myers}\ \emph {et~al.}(2014)\citenamefont {Myers},
  \citenamefont {Das}, \citenamefont {Dartiailh}, \citenamefont {Ohno},
  \citenamefont {Awschalom},\ and\ \citenamefont {Bleszynski~Jayich}}]{MDD+14}%
  \BibitemOpen
  \bibfield  {author} {\bibinfo {author} {\bibfnamefont {B.~A.}\ \bibnamefont
  {Myers}}, \bibinfo {author} {\bibfnamefont {A.}~\bibnamefont {Das}}, \bibinfo
  {author} {\bibfnamefont {M.~C.}\ \bibnamefont {Dartiailh}}, \bibinfo {author}
  {\bibfnamefont {K.}~\bibnamefont {Ohno}}, \bibinfo {author} {\bibfnamefont
  {D.~D.}\ \bibnamefont {Awschalom}}, \ and\ \bibinfo {author} {\bibfnamefont
  {A.~C.}\ \bibnamefont {Bleszynski~Jayich}},\ }\bibfield  {title} {\enquote
  {\bibinfo {title} {Probing {S}urface {N}oise with {D}epth-{C}alibrated
  {S}pins in {D}iamond},}\ }\href@noop {} {\bibfield  {journal} {\bibinfo
  {journal} {Phys.\ Rev.\ Lett.}\ }\textbf {\bibinfo {volume} {113}},\ \bibinfo
  {pages} {027602} (\bibinfo {year} {2014})}\BibitemShut {NoStop}%
\bibitem [{\citenamefont {Loretz}\ \emph {et~al.}(2014)\citenamefont {Loretz},
  \citenamefont {Pezzagna}, \citenamefont {Meijer},\ and\ \citenamefont
  {Degen}}]{LPMD14}%
  \BibitemOpen
  \bibfield  {author} {\bibinfo {author} {\bibfnamefont {M.}~\bibnamefont
  {Loretz}}, \bibinfo {author} {\bibfnamefont {S.}~\bibnamefont {Pezzagna}},
  \bibinfo {author} {\bibfnamefont {J.}~\bibnamefont {Meijer}}, \ and\ \bibinfo
  {author} {\bibfnamefont {C.~L.}\ \bibnamefont {Degen}},\ }\bibfield  {title}
  {\enquote {\bibinfo {title} {Nanoscale nuclear magnetic resonance with a
  1.9-nm-deep nitrogen-vacancy sensor},}\ }\href@noop {} {\bibfield  {journal}
  {\bibinfo  {journal} {Appl.\ Phys.\ Lett.}\ }\textbf {\bibinfo {volume}
  {104}},\ \bibinfo {pages} {033102} (\bibinfo {year} {2014})}\BibitemShut
  {NoStop}%
\bibitem [{\citenamefont {Kim}\ \emph {et~al.}(2014)\citenamefont {Kim},
  \citenamefont {Mamin}, \citenamefont {Sherwood}, \citenamefont {Rettner},
  \citenamefont {Frommer},\ and\ \citenamefont {Rugar}}]{KMS+14}%
  \BibitemOpen
  \bibfield  {author} {\bibinfo {author} {\bibfnamefont {M.}~\bibnamefont
  {Kim}}, \bibinfo {author} {\bibfnamefont {H.~J.}\ \bibnamefont {Mamin}},
  \bibinfo {author} {\bibfnamefont {M.~H.}\ \bibnamefont {Sherwood}}, \bibinfo
  {author} {\bibfnamefont {C.~T.}\ \bibnamefont {Rettner}}, \bibinfo {author}
  {\bibfnamefont {J.}~\bibnamefont {Frommer}}, \ and\ \bibinfo {author}
  {\bibfnamefont {D.}~\bibnamefont {Rugar}},\ }\bibfield  {title} {\enquote
  {\bibinfo {title} {Effect of oxygen plasma and thermal oxidation on shallow
  nitrogen-vacancy centers in diamond},}\ }\href@noop {} {\bibfield  {journal}
  {\bibinfo  {journal} {Appl.\ Phys.\ Lett.}\ }\textbf {\bibinfo {volume}
  {105}},\ \bibinfo {pages} {042406} (\bibinfo {year} {2014})}\BibitemShut
  {NoStop}%
\bibitem [{\citenamefont {Cui}\ \emph {et~al.}(2015)\citenamefont {Cui},
  \citenamefont {Greenspon}, \citenamefont {Ohno}, \citenamefont {Myers},
  \citenamefont {Bleszynski~Jayich}, \citenamefont {Awschalom},\ and\
  \citenamefont {Hu}}]{CGO+15}%
  \BibitemOpen
  \bibfield  {author} {\bibinfo {author} {\bibfnamefont {S.}~\bibnamefont
  {Cui}}, \bibinfo {author} {\bibfnamefont {A.~S.}\ \bibnamefont {Greenspon}},
  \bibinfo {author} {\bibfnamefont {K.}~\bibnamefont {Ohno}}, \bibinfo {author}
  {\bibfnamefont {B.~A.}\ \bibnamefont {Myers}}, \bibinfo {author}
  {\bibfnamefont {A.~C.}\ \bibnamefont {Bleszynski~Jayich}}, \bibinfo {author}
  {\bibfnamefont {D.~D.}\ \bibnamefont {Awschalom}}, \ and\ \bibinfo {author}
  {\bibfnamefont {E.~L.}\ \bibnamefont {Hu}},\ }\bibfield  {title} {\enquote
  {\bibinfo {title} {Reduced {P}lasma-{I}nduced {D}amage to {N}ear-{S}urface
  {N}itrogen-{V}acancy {C}enters in {D}iamond},}\ }\href@noop {} {\bibfield
  {journal} {\bibinfo  {journal} {Nano Lett.}\ }\textbf {\bibinfo {volume}
  {15}},\ \bibinfo {pages} {2887} (\bibinfo {year} {2015})}\BibitemShut
  {NoStop}%
\bibitem [{\citenamefont {de~Oliveira}\ \emph {et~al.}(2015)\citenamefont
  {de~Oliveira}, \citenamefont {Momenzadeh}, \citenamefont {Wang},
  \citenamefont {Konuma}, \citenamefont {Markham}, \citenamefont {Edmonds},
  \citenamefont {Denisenko},\ and\ \citenamefont {Wrachtrup}}]{dOMW+15}%
  \BibitemOpen
  \bibfield  {author} {\bibinfo {author} {\bibfnamefont {F.~F.}\ \bibnamefont
  {de~Oliveira}}, \bibinfo {author} {\bibfnamefont {S.~A.}\ \bibnamefont
  {Momenzadeh}}, \bibinfo {author} {\bibfnamefont {Y.}~\bibnamefont {Wang}},
  \bibinfo {author} {\bibfnamefont {M.}~\bibnamefont {Konuma}}, \bibinfo
  {author} {\bibfnamefont {M.}~\bibnamefont {Markham}}, \bibinfo {author}
  {\bibfnamefont {A.~M.}\ \bibnamefont {Edmonds}}, \bibinfo {author}
  {\bibfnamefont {A.}~\bibnamefont {Denisenko}}, \ and\ \bibinfo {author}
  {\bibfnamefont {J.}~\bibnamefont {Wrachtrup}},\ }\bibfield  {title} {\enquote
  {\bibinfo {title} {Effect of low-damage inductively coupled plasma on shallow
  nitrogen-vacancy centers in diamond},}\ }\href@noop {} {\bibfield  {journal}
  {\bibinfo  {journal} {Appl.\ Phys.\ Lett.}\ }\textbf {\bibinfo {volume}
  {107}},\ \bibinfo {pages} {073107} (\bibinfo {year} {2015})}\BibitemShut
  {NoStop}%
\bibitem [{\citenamefont {Zhang}\ \emph {et~al.}(2017)\citenamefont {Zhang},
  \citenamefont {Zhang}, \citenamefont {Wang}, \citenamefont {Feng},
  \citenamefont {Lin}, \citenamefont {Lou}, \citenamefont {Zhu},\ and\
  \citenamefont {Wang}}]{ZZW+17}%
  \BibitemOpen
  \bibfield  {author} {\bibinfo {author} {\bibfnamefont {W.}~\bibnamefont
  {Zhang}}, \bibinfo {author} {\bibfnamefont {J.}~\bibnamefont {Zhang}},
  \bibinfo {author} {\bibfnamefont {J.}~\bibnamefont {Wang}}, \bibinfo {author}
  {\bibfnamefont {F.}~\bibnamefont {Feng}}, \bibinfo {author} {\bibfnamefont
  {S.}~\bibnamefont {Lin}}, \bibinfo {author} {\bibfnamefont {L.}~\bibnamefont
  {Lou}}, \bibinfo {author} {\bibfnamefont {W.}~\bibnamefont {Zhu}}, \ and\
  \bibinfo {author} {\bibfnamefont {G.}~\bibnamefont {Wang}},\ }\bibfield
  {title} {\enquote {\bibinfo {title} {Depth-dependent decoherence caused by
  surface and external spins for {NV} centers in diamond},}\ }\href@noop {}
  {\bibfield  {journal} {\bibinfo  {journal} {Phys.\ Rev.\ B}\ }\textbf
  {\bibinfo {volume} {96}},\ \bibinfo {pages} {235443} (\bibinfo {year}
  {2017})}\BibitemShut {NoStop}%
\bibitem [{\citenamefont {Ito}\ \emph {et~al.}(2017)\citenamefont {Ito},
  \citenamefont {Saito}, \citenamefont {Sasaki}, \citenamefont {Watanabe},
  \citenamefont {Teraji}, \citenamefont {Itoh},\ and\ \citenamefont
  {Abe}}]{ISS+17}%
  \BibitemOpen
  \bibfield  {author} {\bibinfo {author} {\bibfnamefont {K.}~\bibnamefont
  {Ito}}, \bibinfo {author} {\bibfnamefont {H.}~\bibnamefont {Saito}}, \bibinfo
  {author} {\bibfnamefont {K.}~\bibnamefont {Sasaki}}, \bibinfo {author}
  {\bibfnamefont {H.}~\bibnamefont {Watanabe}}, \bibinfo {author}
  {\bibfnamefont {T.}~\bibnamefont {Teraji}}, \bibinfo {author} {\bibfnamefont
  {K.~M.}\ \bibnamefont {Itoh}}, \ and\ \bibinfo {author} {\bibfnamefont
  {E.}~\bibnamefont {Abe}},\ }\bibfield  {title} {\enquote {\bibinfo {title}
  {Nitrogen-vacancy centers created by {N}$^+$ ion implantation through
  screening {S}i{O}$_2$ layers on diamond},}\ }\href@noop {} {\bibfield
  {journal} {\bibinfo  {journal} {Appl.\ Phys.\ Lett.}\ }\textbf {\bibinfo
  {volume} {110}},\ \bibinfo {pages} {213105} (\bibinfo {year}
  {2017})}\BibitemShut {NoStop}%
\bibitem [{\citenamefont {Abe}\ and\ \citenamefont {Sasaki}(2018)}]{AS18}%
  \BibitemOpen
  \bibfield  {author} {\bibinfo {author} {\bibfnamefont {E.}~\bibnamefont
  {Abe}}\ and\ \bibinfo {author} {\bibfnamefont {K.}~\bibnamefont {Sasaki}},\
  }\bibfield  {title} {\enquote {\bibinfo {title} {Tutorial: {M}agnetic
  resonance with nitrogen-vacancy centers in diamond---microwave engineering,
  materials science, and magnetometry},}\ }\href@noop {} {\bibfield  {journal}
  {\bibinfo  {journal} {J.\ Appl.\ Phys.}\ }\textbf {\bibinfo {volume} {123}},\
  \bibinfo {pages} {161101} (\bibinfo {year} {2018})}\BibitemShut {NoStop}%
\bibitem [{\citenamefont {Sasaki}, \citenamefont {Itoh},\ and\ \citenamefont
  {Abe}(2018)}]{SIA18}%
  \BibitemOpen
  \bibfield  {author} {\bibinfo {author} {\bibfnamefont {K.}~\bibnamefont
  {Sasaki}}, \bibinfo {author} {\bibfnamefont {K.~M.}\ \bibnamefont {Itoh}}, \
  and\ \bibinfo {author} {\bibfnamefont {E.}~\bibnamefont {Abe}},\ }\bibfield
  {title} {\enquote {\bibinfo {title} {Determination of the position of a
  single nuclear spin from free nuclear precessions detected by a solid-state
  quantum sensor},}\ }\href@noop {} {\bibfield  {journal} {\bibinfo  {journal}
  {Phys.\ Rev.\ B}\ }\textbf {\bibinfo {volume} {98}},\ \bibinfo {pages}
  {121405} (\bibinfo {year} {2018})}\BibitemShut {NoStop}%
\bibitem [{\citenamefont {Misonou}\ \emph {et~al.}(2020)\citenamefont
  {Misonou}, \citenamefont {Sasaki}, \citenamefont {Ishizu}, \citenamefont
  {Monnai}, \citenamefont {Itoh},\ and\ \citenamefont {Abe}}]{MSI+20}%
  \BibitemOpen
  \bibfield  {author} {\bibinfo {author} {\bibfnamefont {D.}~\bibnamefont
  {Misonou}}, \bibinfo {author} {\bibfnamefont {K.}~\bibnamefont {Sasaki}},
  \bibinfo {author} {\bibfnamefont {S.}~\bibnamefont {Ishizu}}, \bibinfo
  {author} {\bibfnamefont {Y.}~\bibnamefont {Monnai}}, \bibinfo {author}
  {\bibfnamefont {K.~M.}\ \bibnamefont {Itoh}}, \ and\ \bibinfo {author}
  {\bibfnamefont {E.}~\bibnamefont {Abe}},\ }\bibfield  {title} {\enquote
  {\bibinfo {title} {Construction and operation of a tabletop system for
  nanoscale magnetometry with single nitrogen-vacancy centers in diamond},}\
  }\href@noop {} {\bibfield  {journal} {\bibinfo  {journal} {AIP Adv.}\
  }\textbf {\bibinfo {volume} {10}},\ \bibinfo {pages} {025206} (\bibinfo
  {year} {2020})}\BibitemShut {NoStop}%
\bibitem [{\citenamefont {Osterkamp}\ \emph {et~al.}(2013)\citenamefont
  {Osterkamp}, \citenamefont {Scharpf}, \citenamefont {Pezzagna}, \citenamefont
  {Meijer}, \citenamefont {Diemant}, \citenamefont {Behm}, \citenamefont
  {Naydenov},\ and\ \citenamefont {Jelezko}}]{OSP+13}%
  \BibitemOpen
  \bibfield  {author} {\bibinfo {author} {\bibfnamefont {C.}~\bibnamefont
  {Osterkamp}}, \bibinfo {author} {\bibfnamefont {J.}~\bibnamefont {Scharpf}},
  \bibinfo {author} {\bibfnamefont {S.}~\bibnamefont {Pezzagna}}, \bibinfo
  {author} {\bibfnamefont {J.}~\bibnamefont {Meijer}}, \bibinfo {author}
  {\bibfnamefont {T.}~\bibnamefont {Diemant}}, \bibinfo {author} {\bibfnamefont
  {R.~J.}\ \bibnamefont {Behm}}, \bibinfo {author} {\bibfnamefont
  {B.}~\bibnamefont {Naydenov}}, \ and\ \bibinfo {author} {\bibfnamefont
  {F.}~\bibnamefont {Jelezko}},\ }\bibfield  {title} {\enquote {\bibinfo
  {title} {Increasing the creation yield of shallow single defects in diamond
  by surface plasma treatment},}\ }\href@noop {} {\bibfield  {journal}
  {\bibinfo  {journal} {Appl.\ Phys.\ Lett.}\ }\textbf {\bibinfo {volume}
  {103}},\ \bibinfo {pages} {193118} (\bibinfo {year} {2013})}\BibitemShut
  {NoStop}%
\bibitem [{\citenamefont {Yamamoto}\ \emph {et~al.}(2013)\citenamefont
  {Yamamoto}, \citenamefont {Umeda}, \citenamefont {Watanabe}, \citenamefont
  {Onoda}, \citenamefont {Markham}, \citenamefont {Twitchen}, \citenamefont
  {Naydenov}, \citenamefont {McGuinness}, \citenamefont {Teraji}, \citenamefont
  {Koizumi}, \citenamefont {Dolde}, \citenamefont {Fedder}, \citenamefont
  {Honert}, \citenamefont {Wrachtrup}, \citenamefont {Ohshima}, \citenamefont
  {Jelezko},\ and\ \citenamefont {Isoya}}]{YUW+13}%
  \BibitemOpen
  \bibfield  {author} {\bibinfo {author} {\bibfnamefont {T.}~\bibnamefont
  {Yamamoto}}, \bibinfo {author} {\bibfnamefont {T.}~\bibnamefont {Umeda}},
  \bibinfo {author} {\bibfnamefont {K.}~\bibnamefont {Watanabe}}, \bibinfo
  {author} {\bibfnamefont {S.}~\bibnamefont {Onoda}}, \bibinfo {author}
  {\bibfnamefont {M.~L.}\ \bibnamefont {Markham}}, \bibinfo {author}
  {\bibfnamefont {D.~J.}\ \bibnamefont {Twitchen}}, \bibinfo {author}
  {\bibfnamefont {B.}~\bibnamefont {Naydenov}}, \bibinfo {author}
  {\bibfnamefont {L.~P.}\ \bibnamefont {McGuinness}}, \bibinfo {author}
  {\bibfnamefont {T.}~\bibnamefont {Teraji}}, \bibinfo {author} {\bibfnamefont
  {S.}~\bibnamefont {Koizumi}}, \bibinfo {author} {\bibfnamefont
  {F.}~\bibnamefont {Dolde}}, \bibinfo {author} {\bibfnamefont
  {H.}~\bibnamefont {Fedder}}, \bibinfo {author} {\bibfnamefont
  {J.}~\bibnamefont {Honert}}, \bibinfo {author} {\bibfnamefont
  {J.}~\bibnamefont {Wrachtrup}}, \bibinfo {author} {\bibfnamefont
  {T.}~\bibnamefont {Ohshima}}, \bibinfo {author} {\bibfnamefont
  {F.}~\bibnamefont {Jelezko}}, \ and\ \bibinfo {author} {\bibfnamefont
  {J.}~\bibnamefont {Isoya}},\ }\bibfield  {title} {\enquote {\bibinfo {title}
  {Extending spin coherence times of diamond qubits by high-temperature
  annealing},}\ }\href@noop {} {\bibfield  {journal} {\bibinfo  {journal}
  {Phys.\ Rev.\ B}\ }\textbf {\bibinfo {volume} {88}},\ \bibinfo {pages}
  {075206} (\bibinfo {year} {2013})}\BibitemShut {NoStop}%
\bibitem [{\citenamefont {Antonov}\ \emph {et~al.}(2014)\citenamefont
  {Antonov}, \citenamefont {H{\"a}usermann}, \citenamefont {Aird},
  \citenamefont {Roth}, \citenamefont {Trebin}, \citenamefont {M{\"u}ller},
  \citenamefont {McGuinness}, \citenamefont {Jelezko}, \citenamefont
  {Yamamoto}, \citenamefont {Isoya}, \citenamefont {Pezzagna}, \citenamefont
  {Meijer}, ,\ and\ \citenamefont {Wrachtrup}}]{AHA+14}%
  \BibitemOpen
  \bibfield  {author} {\bibinfo {author} {\bibfnamefont {D.}~\bibnamefont
  {Antonov}}, \bibinfo {author} {\bibfnamefont {T.}~\bibnamefont
  {H{\"a}usermann}}, \bibinfo {author} {\bibfnamefont {A.}~\bibnamefont
  {Aird}}, \bibinfo {author} {\bibfnamefont {J.}~\bibnamefont {Roth}}, \bibinfo
  {author} {\bibfnamefont {H.-R.}\ \bibnamefont {Trebin}}, \bibinfo {author}
  {\bibfnamefont {C.}~\bibnamefont {M{\"u}ller}}, \bibinfo {author}
  {\bibfnamefont {L.}~\bibnamefont {McGuinness}}, \bibinfo {author}
  {\bibfnamefont {F.}~\bibnamefont {Jelezko}}, \bibinfo {author} {\bibfnamefont
  {T.}~\bibnamefont {Yamamoto}}, \bibinfo {author} {\bibfnamefont
  {J.}~\bibnamefont {Isoya}}, \bibinfo {author} {\bibfnamefont
  {S.}~\bibnamefont {Pezzagna}}, \bibinfo {author} {\bibfnamefont
  {J.}~\bibnamefont {Meijer}}, , \ and\ \bibinfo {author} {\bibfnamefont
  {J.}~\bibnamefont {Wrachtrup}},\ }\bibfield  {title} {\enquote {\bibinfo
  {title} {Statistical investigations on nitrogen-vacancy center creation},}\
  }\href@noop {} {\bibfield  {journal} {\bibinfo  {journal} {Appl.\ Phys.\
  Lett.}\ }\textbf {\bibinfo {volume} {104}},\ \bibinfo {pages} {012105}
  (\bibinfo {year} {2014})}\BibitemShut {NoStop}%
\bibitem [{\citenamefont {Hahn}(1950)}]{H50}%
  \BibitemOpen
  \bibfield  {author} {\bibinfo {author} {\bibfnamefont {E.~L.}\ \bibnamefont
  {Hahn}},\ }\bibfield  {title} {\enquote {\bibinfo {title} {Spin {E}choes},}\
  }\href@noop {} {\bibfield  {journal} {\bibinfo  {journal} {Phys.\ Rev.}\
  }\textbf {\bibinfo {volume} {80}},\ \bibinfo {pages} {580} (\bibinfo {year}
  {1950})}\BibitemShut {NoStop}%
\bibitem [{\citenamefont {Gullion}, \citenamefont {Baker},\ and\ \citenamefont
  {Conradi}(1990)}]{GBC90}%
  \BibitemOpen
  \bibfield  {author} {\bibinfo {author} {\bibfnamefont {T.}~\bibnamefont
  {Gullion}}, \bibinfo {author} {\bibfnamefont {D.~B.}\ \bibnamefont {Baker}},
  \ and\ \bibinfo {author} {\bibfnamefont {M.~S.}\ \bibnamefont {Conradi}},\
  }\bibfield  {title} {\enquote {\bibinfo {title} {New, compensated
  {C}arr-{P}urcell sequences},}\ }\href@noop {} {\bibfield  {journal} {\bibinfo
   {journal} {J.\ Mag.\ Res.}\ }\textbf {\bibinfo {volume} {89}},\ \bibinfo
  {pages} {479} (\bibinfo {year} {1990})}\BibitemShut {NoStop}%
\bibitem [{\citenamefont {Pham}\ \emph {et~al.}(2016)\citenamefont {Pham},
  \citenamefont {DeVience}, \citenamefont {Casola}, \citenamefont {Lovchinsky},
  \citenamefont {Sushkov}, \citenamefont {Bersin}, \citenamefont {Lee},
  \citenamefont {Urbach}, \citenamefont {Cappellaro}, \citenamefont {Park},
  \citenamefont {Yacoby}, \citenamefont {Lukin},\ and\ \citenamefont
  {Walsworth}}]{PDC+16}%
  \BibitemOpen
  \bibfield  {author} {\bibinfo {author} {\bibfnamefont {L.~M.}\ \bibnamefont
  {Pham}}, \bibinfo {author} {\bibfnamefont {S.~J.}\ \bibnamefont {DeVience}},
  \bibinfo {author} {\bibfnamefont {F.}~\bibnamefont {Casola}}, \bibinfo
  {author} {\bibfnamefont {I.}~\bibnamefont {Lovchinsky}}, \bibinfo {author}
  {\bibfnamefont {A.~O.}\ \bibnamefont {Sushkov}}, \bibinfo {author}
  {\bibfnamefont {E.}~\bibnamefont {Bersin}}, \bibinfo {author} {\bibfnamefont
  {J.}~\bibnamefont {Lee}}, \bibinfo {author} {\bibfnamefont {E.}~\bibnamefont
  {Urbach}}, \bibinfo {author} {\bibfnamefont {P.}~\bibnamefont {Cappellaro}},
  \bibinfo {author} {\bibfnamefont {H.}~\bibnamefont {Park}}, \bibinfo {author}
  {\bibfnamefont {A.}~\bibnamefont {Yacoby}}, \bibinfo {author} {\bibfnamefont
  {M.}~\bibnamefont {Lukin}}, \ and\ \bibinfo {author} {\bibfnamefont {R.~L.}\
  \bibnamefont {Walsworth}},\ }\bibfield  {title} {\enquote {\bibinfo {title}
  {{NMR} technique for determining the depth of shallow nitrogen-vacancy
  centers in diamond},}\ }\href@noop {} {\bibfield  {journal} {\bibinfo
  {journal} {Phys.\ Rev.\ B}\ }\textbf {\bibinfo {volume} {93}},\ \bibinfo
  {pages} {045425} (\bibinfo {year} {2016})}\BibitemShut {NoStop}%
\bibitem [{\citenamefont {Rosskopf}\ \emph {et~al.}(2014)\citenamefont
  {Rosskopf}, \citenamefont {Dussaux}, \citenamefont {Ohashi}, \citenamefont
  {Loretz}, \citenamefont {Schirhagl}, \citenamefont {Watanabe}, \citenamefont
  {Shikata}, \citenamefont {Itoh},\ and\ \citenamefont {Degen}}]{RDO+14}%
  \BibitemOpen
  \bibfield  {author} {\bibinfo {author} {\bibfnamefont {T.}~\bibnamefont
  {Rosskopf}}, \bibinfo {author} {\bibfnamefont {A.}~\bibnamefont {Dussaux}},
  \bibinfo {author} {\bibfnamefont {K.}~\bibnamefont {Ohashi}}, \bibinfo
  {author} {\bibfnamefont {M.}~\bibnamefont {Loretz}}, \bibinfo {author}
  {\bibfnamefont {R.}~\bibnamefont {Schirhagl}}, \bibinfo {author}
  {\bibfnamefont {H.}~\bibnamefont {Watanabe}}, \bibinfo {author}
  {\bibfnamefont {S.}~\bibnamefont {Shikata}}, \bibinfo {author} {\bibfnamefont
  {K.~M.}\ \bibnamefont {Itoh}}, \ and\ \bibinfo {author} {\bibfnamefont
  {C.~L.}\ \bibnamefont {Degen}},\ }\bibfield  {title} {\enquote {\bibinfo
  {title} {Investigation of {S}urface {M}agnetic {N}oise by {S}hallow {S}pins
  in {D}iamond},}\ }\href@noop {} {\bibfield  {journal} {\bibinfo  {journal}
  {Phys.\ Rev.\ Lett.}\ }\textbf {\bibinfo {volume} {112}},\ \bibinfo {pages}
  {147602} (\bibinfo {year} {2014})}\BibitemShut {NoStop}%
\bibitem [{\citenamefont {Bar-Gill}\ \emph {et~al.}(2012)\citenamefont
  {Bar-Gill}, \citenamefont {Pham}, \citenamefont {Belthangady}, \citenamefont
  {Le~Sage}, \citenamefont {Cappellaro}, \citenamefont {Maze},\ and\
  \citenamefont {Lukin}}]{BPB+12}%
  \BibitemOpen
  \bibfield  {author} {\bibinfo {author} {\bibfnamefont {N.}~\bibnamefont
  {Bar-Gill}}, \bibinfo {author} {\bibfnamefont {L.~M.}\ \bibnamefont {Pham}},
  \bibinfo {author} {\bibfnamefont {C.}~\bibnamefont {Belthangady}}, \bibinfo
  {author} {\bibfnamefont {D.}~\bibnamefont {Le~Sage}}, \bibinfo {author}
  {\bibfnamefont {P.}~\bibnamefont {Cappellaro}}, \bibinfo {author}
  {\bibfnamefont {J.~R.}\ \bibnamefont {Maze}}, \ and\ \bibinfo {author}
  {\bibfnamefont {M.~D.}\ \bibnamefont {Lukin}},\ }\bibfield  {title} {\enquote
  {\bibinfo {title} {Suppression of spin-bath dynamics for improved coherence
  of multi-spin-qubit systems},}\ }\href@noop {} {\bibfield  {journal}
  {\bibinfo  {journal} {Nat.\ Commun.}\ }\textbf {\bibinfo {volume} {3}},\
  \bibinfo {pages} {858} (\bibinfo {year} {2012})}\BibitemShut {NoStop}%
\bibitem [{\citenamefont {Romach}\ \emph {et~al.}(2015)\citenamefont {Romach},
  \citenamefont {M{\"u}ller}, \citenamefont {Unden}, \citenamefont {Rogers},
  \citenamefont {Isoda}, \citenamefont {Itoh}, \citenamefont {Markham},
  \citenamefont {Stacey}, \citenamefont {Meijer}, \citenamefont {Pezzagna},
  \citenamefont {Naydenov}, \citenamefont {McGuinness}, \citenamefont
  {Bar-Gill},\ and\ \citenamefont {Jelezko}}]{RMU+15}%
  \BibitemOpen
  \bibfield  {author} {\bibinfo {author} {\bibfnamefont {Y.}~\bibnamefont
  {Romach}}, \bibinfo {author} {\bibfnamefont {C.}~\bibnamefont {M{\"u}ller}},
  \bibinfo {author} {\bibfnamefont {T.}~\bibnamefont {Unden}}, \bibinfo
  {author} {\bibfnamefont {L.~J.}\ \bibnamefont {Rogers}}, \bibinfo {author}
  {\bibfnamefont {T.}~\bibnamefont {Isoda}}, \bibinfo {author} {\bibfnamefont
  {K.~M.}\ \bibnamefont {Itoh}}, \bibinfo {author} {\bibfnamefont
  {M.}~\bibnamefont {Markham}}, \bibinfo {author} {\bibfnamefont
  {A.}~\bibnamefont {Stacey}}, \bibinfo {author} {\bibfnamefont
  {J.}~\bibnamefont {Meijer}}, \bibinfo {author} {\bibfnamefont
  {S.}~\bibnamefont {Pezzagna}}, \bibinfo {author} {\bibfnamefont
  {B.}~\bibnamefont {Naydenov}}, \bibinfo {author} {\bibfnamefont {L.~P.}\
  \bibnamefont {McGuinness}}, \bibinfo {author} {\bibfnamefont
  {N.}~\bibnamefont {Bar-Gill}}, \ and\ \bibinfo {author} {\bibfnamefont
  {F.}~\bibnamefont {Jelezko}},\ }\bibfield  {title} {\enquote {\bibinfo
  {title} {Spectroscopy of {S}urface-{I}nduced {N}oise {U}sing {S}hallow
  {S}pins in {D}iamond},}\ }\href@noop {} {\bibfield  {journal} {\bibinfo
  {journal} {Phys.\ Rev.\ Lett.}\ }\textbf {\bibinfo {volume} {114}},\ \bibinfo
  {pages} {017601} (\bibinfo {year} {2015})}\BibitemShut {NoStop}%
\bibitem [{\citenamefont {Myers}, \citenamefont {Ariyaratne},\ and\
  \citenamefont {Bleszynski~Jayich}(2017)}]{MAJ17}%
  \BibitemOpen
  \bibfield  {author} {\bibinfo {author} {\bibfnamefont {B.~A.}\ \bibnamefont
  {Myers}}, \bibinfo {author} {\bibfnamefont {A.}~\bibnamefont {Ariyaratne}}, \
  and\ \bibinfo {author} {\bibfnamefont {A.~C.}\ \bibnamefont
  {Bleszynski~Jayich}},\ }\bibfield  {title} {\enquote {\bibinfo {title}
  {Double-{Q}uantum {S}pin-{R}elaxation {L}imits to {C}oherence of
  {N}ear-{S}urface {N}itrogen-{V}acancy {C}enters},}\ }\href@noop {} {\bibfield
   {journal} {\bibinfo  {journal} {Phys.\ Rev.\ Lett.}\ }\textbf {\bibinfo
  {volume} {118}},\ \bibinfo {pages} {197201} (\bibinfo {year}
  {2017})}\BibitemShut {NoStop}%
\bibitem [{\citenamefont {Gardill}, \citenamefont {Cambria},\ and\
  \citenamefont {Kolkowitz}(2020)}]{GCK20}%
  \BibitemOpen
  \bibfield  {author} {\bibinfo {author} {\bibfnamefont {A.}~\bibnamefont
  {Gardill}}, \bibinfo {author} {\bibfnamefont {M.}~\bibnamefont {Cambria}}, \
  and\ \bibinfo {author} {\bibfnamefont {S.}~\bibnamefont {Kolkowitz}},\
  }\bibfield  {title} {\enquote {\bibinfo {title} {Fast {R}elaxation on
  {Q}utrit {T}ransitions of {N}itrogen-{V}acancy {C}enters in
  {N}anodiamonds},}\ }\href@noop {} {\bibfield  {journal} {\bibinfo  {journal}
  {Phys.\ Rev.\ Appl.}\ }\textbf {\bibinfo {volume} {10}},\ \bibinfo {pages}
  {034010} (\bibinfo {year} {2020})}\BibitemShut {NoStop}%
\bibitem [{\citenamefont {Baldwin}\ \emph {et~al.}(2014)\citenamefont
  {Baldwin}, \citenamefont {Downes}, \citenamefont {McMahon}, \citenamefont
  {Bradac},\ and\ \citenamefont {Mildren}}]{BDM+14}%
  \BibitemOpen
  \bibfield  {author} {\bibinfo {author} {\bibfnamefont {C.~G.}\ \bibnamefont
  {Baldwin}}, \bibinfo {author} {\bibfnamefont {J.~E.}\ \bibnamefont {Downes}},
  \bibinfo {author} {\bibfnamefont {C.~J.}\ \bibnamefont {McMahon}}, \bibinfo
  {author} {\bibfnamefont {C.}~\bibnamefont {Bradac}}, \ and\ \bibinfo {author}
  {\bibfnamefont {R.~P.}\ \bibnamefont {Mildren}},\ }\bibfield  {title}
  {\enquote {\bibinfo {title} {Nanostructuring and oxidation of diamond by
  two-photon ultraviolet surface excitation: {A}n {XPS} and {NEXAFS} study},}\
  }\href@noop {} {\bibfield  {journal} {\bibinfo  {journal} {Phys.\ Rev.\ B}\
  }\textbf {\bibinfo {volume} {89}},\ \bibinfo {pages} {195422} (\bibinfo
  {year} {2014})}\BibitemShut {NoStop}%
\bibitem [{\citenamefont {Zeigler}()}]{SRIM}%
  \BibitemOpen
  \bibfield  {author} {\bibinfo {author} {\bibfnamefont {J.~F.}\ \bibnamefont
  {Zeigler}},\ }\href@noop {} {\emph {\bibinfo {title} {The {S}topping and
  {R}ange of {I}ons in {M}atter, {SRIM}-2013}}},\ \bibinfo {note}
  {http://www.srim.org/}\BibitemShut {NoStop}%
\bibitem [{\citenamefont {Zeigler}(2004)}]{Z04}%
  \BibitemOpen
  \bibfield  {author} {\bibinfo {author} {\bibfnamefont {J.~F.}\ \bibnamefont
  {Zeigler}},\ }\bibfield  {title} {\enquote {\bibinfo {title} {S{RIM}-2003},}\
  }\href@noop {} {\bibfield  {journal} {\bibinfo  {journal} {Nucl.\ Instrum.\
  Methods Phys.\ Res.\ B}\ }\textbf {\bibinfo {volume} {219-220}},\ \bibinfo
  {pages} {1027} (\bibinfo {year} {2004})}\BibitemShut {NoStop}%
\bibitem [{\citenamefont {Teraji}(2015)}]{T15}%
  \BibitemOpen
  \bibfield  {author} {\bibinfo {author} {\bibfnamefont {T.}~\bibnamefont
  {Teraji}},\ }\bibfield  {title} {\enquote {\bibinfo {title} {High-quality and
  high-purity homoepitaxial diamond (100) film growth under high oxygen
  concentration condition},}\ }\href@noop {} {\bibfield  {journal} {\bibinfo
  {journal} {J.\ Appl.\ Phys.}\ }\textbf {\bibinfo {volume} {118}},\ \bibinfo
  {pages} {115304} (\bibinfo {year} {2015})}\BibitemShut {NoStop}%
\bibitem [{\citenamefont {Childress}\ \emph {et~al.}(2006)\citenamefont
  {Childress}, \citenamefont {Gurudev~Dutt}, \citenamefont {Taylor},
  \citenamefont {Zibrov}, \citenamefont {Jelezko}, \citenamefont {Wrachtrup},
  \citenamefont {Hemmer},\ and\ \citenamefont {Lukin}}]{CDT+06}%
  \BibitemOpen
  \bibfield  {author} {\bibinfo {author} {\bibfnamefont {L.}~\bibnamefont
  {Childress}}, \bibinfo {author} {\bibfnamefont {M.~V.}\ \bibnamefont
  {Gurudev~Dutt}}, \bibinfo {author} {\bibfnamefont {J.~M.}\ \bibnamefont
  {Taylor}}, \bibinfo {author} {\bibfnamefont {A.~S.}\ \bibnamefont {Zibrov}},
  \bibinfo {author} {\bibfnamefont {F.}~\bibnamefont {Jelezko}}, \bibinfo
  {author} {\bibfnamefont {J.}~\bibnamefont {Wrachtrup}}, \bibinfo {author}
  {\bibfnamefont {P.~R.}\ \bibnamefont {Hemmer}}, \ and\ \bibinfo {author}
  {\bibfnamefont {M.~D.}\ \bibnamefont {Lukin}},\ }\bibfield  {title} {\enquote
  {\bibinfo {title} {Coherent {D}ynamics of {C}oupled {E}lectron and {N}uclear
  {S}pin {Q}ubits in {D}iamond},}\ }\href@noop {} {\bibfield  {journal}
  {\bibinfo  {journal} {Science}\ }\textbf {\bibinfo {volume} {314}},\ \bibinfo
  {pages} {281} (\bibinfo {year} {2006})}\BibitemShut {NoStop}%
\bibitem [{\citenamefont {Loretz}\ \emph {et~al.}(2015)\citenamefont {Loretz},
  \citenamefont {Boss}, \citenamefont {Rosskopf}, \citenamefont {Mamin},
  \citenamefont {Rugar},\ and\ \citenamefont {Degen}}]{LBR+15}%
  \BibitemOpen
  \bibfield  {author} {\bibinfo {author} {\bibfnamefont {M.}~\bibnamefont
  {Loretz}}, \bibinfo {author} {\bibfnamefont {J.~M.}\ \bibnamefont {Boss}},
  \bibinfo {author} {\bibfnamefont {T.}~\bibnamefont {Rosskopf}}, \bibinfo
  {author} {\bibfnamefont {H.~J.}\ \bibnamefont {Mamin}}, \bibinfo {author}
  {\bibfnamefont {D.}~\bibnamefont {Rugar}}, \ and\ \bibinfo {author}
  {\bibfnamefont {C.~L.}\ \bibnamefont {Degen}},\ }\bibfield  {title} {\enquote
  {\bibinfo {title} {Spurious {H}armonic {R}esponse of {M}ultipulse {Q}uantum
  {S}ensing {S}equences},}\ }\href@noop {} {\bibfield  {journal} {\bibinfo
  {journal} {Phys.\ Rev.\ X}\ }\textbf {\bibinfo {volume} {5}},\ \bibinfo
  {pages} {021009} (\bibinfo {year} {2015})}\BibitemShut {NoStop}%
\bibitem [{\citenamefont {Fu}\ \emph {et~al.}(2010)\citenamefont {Fu},
  \citenamefont {Santori}, \citenamefont {Barclay},\ and\ \citenamefont
  {Beausoleil}}]{FSB+10}%
  \BibitemOpen
  \bibfield  {author} {\bibinfo {author} {\bibfnamefont {K.-M.~C.}\
  \bibnamefont {Fu}}, \bibinfo {author} {\bibfnamefont {C.}~\bibnamefont
  {Santori}}, \bibinfo {author} {\bibfnamefont {P.~E.}\ \bibnamefont
  {Barclay}}, \ and\ \bibinfo {author} {\bibfnamefont {R.~G.}\ \bibnamefont
  {Beausoleil}},\ }\bibfield  {title} {\enquote {\bibinfo {title} {Conversion
  of neutral nitrogen-vacancy centers to negatively charged nitrogen-vacancy
  centers through selective oxidation},}\ }\href@noop {} {\bibfield  {journal}
  {\bibinfo  {journal} {Appl.\ Phys.\ Lett.}\ }\textbf {\bibinfo {volume}
  {96}},\ \bibinfo {pages} {121907} (\bibinfo {year} {2010})}\BibitemShut
  {NoStop}%
\bibitem [{\citenamefont {Sasaki}\ \emph {et~al.}(2016)\citenamefont {Sasaki},
  \citenamefont {Monnai}, \citenamefont {Saijo}, \citenamefont {Fujita},
  \citenamefont {Watanabe}, \citenamefont {Ishi-Hayase}, \citenamefont {Itoh},\
  and\ \citenamefont {Abe}}]{SMS+16}%
  \BibitemOpen
  \bibfield  {author} {\bibinfo {author} {\bibfnamefont {K.}~\bibnamefont
  {Sasaki}}, \bibinfo {author} {\bibfnamefont {Y.}~\bibnamefont {Monnai}},
  \bibinfo {author} {\bibfnamefont {S.}~\bibnamefont {Saijo}}, \bibinfo
  {author} {\bibfnamefont {R.}~\bibnamefont {Fujita}}, \bibinfo {author}
  {\bibfnamefont {H.}~\bibnamefont {Watanabe}}, \bibinfo {author}
  {\bibfnamefont {J.}~\bibnamefont {Ishi-Hayase}}, \bibinfo {author}
  {\bibfnamefont {K.~M.}\ \bibnamefont {Itoh}}, \ and\ \bibinfo {author}
  {\bibfnamefont {E.}~\bibnamefont {Abe}},\ }\bibfield  {title} {\enquote
  {\bibinfo {title} {Broadband, large-area microwave antenna for optically
  detected magnetic resonance of nitrogen-vacancy centers in diamond},}\
  }\href@noop {} {\bibfield  {journal} {\bibinfo  {journal} {Rev.\ Sci.\
  Instrum.}\ }\textbf {\bibinfo {volume} {87}},\ \bibinfo {pages} {053904}
  (\bibinfo {year} {2016})}\BibitemShut {NoStop}%
\bibitem [{\citenamefont {Schweiger}\ and\ \citenamefont
  {Jeschke}(2001)}]{SJ01}%
  \BibitemOpen
  \bibfield  {author} {\bibinfo {author} {\bibfnamefont {A.}~\bibnamefont
  {Schweiger}}\ and\ \bibinfo {author} {\bibfnamefont {G.}~\bibnamefont
  {Jeschke}},\ }\href@noop {} {\emph {\bibinfo {title} {Principles of {P}ulse
  {E}lectron {P}aramagnetic {R}esonance}}}\ (\bibinfo  {publisher} {Oxford
  University Press, Oxford},\ \bibinfo {year} {2001})\BibitemShut {NoStop}%
\bibitem [{\citenamefont {Sasaki}\ \emph {et~al.}(2017)\citenamefont {Sasaki},
  \citenamefont {Kleinsasser}, \citenamefont {Zhu}, \citenamefont {Li},
  \citenamefont {Watanabe}, \citenamefont {Fu}, \citenamefont {Itoh},\ and\
  \citenamefont {Abe}}]{SKZ+17}%
  \BibitemOpen
  \bibfield  {author} {\bibinfo {author} {\bibfnamefont {K.}~\bibnamefont
  {Sasaki}}, \bibinfo {author} {\bibfnamefont {E.~E.}\ \bibnamefont
  {Kleinsasser}}, \bibinfo {author} {\bibfnamefont {Z.}~\bibnamefont {Zhu}},
  \bibinfo {author} {\bibfnamefont {W.-D.}\ \bibnamefont {Li}}, \bibinfo
  {author} {\bibfnamefont {H.}~\bibnamefont {Watanabe}}, \bibinfo {author}
  {\bibfnamefont {K.-M.~C.}\ \bibnamefont {Fu}}, \bibinfo {author}
  {\bibfnamefont {K.~M.}\ \bibnamefont {Itoh}}, \ and\ \bibinfo {author}
  {\bibfnamefont {E.}~\bibnamefont {Abe}},\ }\bibfield  {title} {\enquote
  {\bibinfo {title} {Dynamic nuclear polarization enhanced magnetic field
  sensitivity and decoherence spectroscopy of an ensemble of near-surface
  nitrogen-vacancy centers in diamond},}\ }\href@noop {} {\bibfield  {journal}
  {\bibinfo  {journal} {Appl.\ Phys.\ Lett.}\ }\textbf {\bibinfo {volume}
  {110}},\ \bibinfo {pages} {192407} (\bibinfo {year} {2017})}\BibitemShut
  {NoStop}%
\bibitem [{\citenamefont {Bluvstein}\ \emph {et~al.}(2019)\citenamefont
  {Bluvstein}, \citenamefont {Zhang}, \citenamefont {McLellan}, \citenamefont
  {Williams},\ and\ \citenamefont {Bleszynski~Jayich}}]{BZM+19}%
  \BibitemOpen
  \bibfield  {author} {\bibinfo {author} {\bibfnamefont {D.}~\bibnamefont
  {Bluvstein}}, \bibinfo {author} {\bibfnamefont {Z.}~\bibnamefont {Zhang}},
  \bibinfo {author} {\bibfnamefont {C.~A.}\ \bibnamefont {McLellan}}, \bibinfo
  {author} {\bibfnamefont {N.~R.}\ \bibnamefont {Williams}}, \ and\ \bibinfo
  {author} {\bibfnamefont {A.~C.}\ \bibnamefont {Bleszynski~Jayich}},\
  }\bibfield  {title} {\enquote {\bibinfo {title} {Extending the {Q}uantum
  {C}oherence of a {N}ear-{S}urface {Q}ubit by {C}oherently {D}riving the
  {P}aramagnetic {S}urface {E}nvironment},}\ }\href@noop {} {\bibfield
  {journal} {\bibinfo  {journal} {Phys.\ Rev.\ Lett.}\ }\textbf {\bibinfo
  {volume} {123}},\ \bibinfo {pages} {146804} (\bibinfo {year}
  {2019})}\BibitemShut {NoStop}%
\end{thebibliography}%
\end{document}